\documentclass[twocolumn,showpacs,prx,superscriptaddress,floatfix]{revtex4}

\usepackage{color}
\usepackage{tabularx}
\usepackage{epsfig}
\usepackage{amsmath}
\usepackage{amssymb}
\usepackage{graphicx}
\usepackage{wasysym}
\usepackage{dcolumn}
\usepackage{bm}
\usepackage{subfigure}
\usepackage{psfrag}
\usepackage{subfigure}
\usepackage{times}

\begin{document}

\title{Seeking Quantum Speedup through Spin Glasses:\\ 
The Good, the Bad, and the Ugly}

\author{Helmut G.~Katzgraber}
\affiliation{Department of Physics and Astronomy, Texas A\&M University,
College Station, Texas 77843-4242, USA}
\affiliation{Materials Science and Engineering Program, Texas A\&M
University, College Station, Texas 77843, USA}
\affiliation{Santa Fe Institute, 1399 Hyde Park Road, Santa Fe, New Mexico 
87501, USA}

\author{Firas Hamze}
\affiliation {D-Wave Systems, Inc., 3033 Beta Avenue, Burnaby, British
Columbia, V5G 4M9, Canada}

\author{Zheng Zhu}
\affiliation{Department of Physics and Astronomy, Texas A\&M University,
College Station, Texas 77843-4242, USA}

\author{Andrew J.~Ochoa}
\affiliation{Department of Physics and Astronomy, Texas A\&M University,
College Station, Texas 77843-4242, USA}

\author{H.~Munoz-Bauza}
\affiliation{Department of Physics and Astronomy, Texas A\&M University,
College Station, Texas 77843-4242, USA}

\date{\today}

\begin{abstract}

There has been considerable progress in the design and construction of
quantum annealing devices. However, a conclusive detection of quantum
speedup over traditional silicon-based machines remains elusive, despite
multiple careful studies. In this work we outline strategies to design
hard tunable benchmark instances based on insights from the study of
spin glasses---the archetypal random benchmark problem for novel
algorithms and optimization devices. We propose to complement
head-to-head scaling studies that compare quantum annealing machines to
state-of-the-art classical codes with an approach that compares the
performance of different algorithms and/or computing architectures on
different classes of computationally hard tunable spin-glass instances.
The advantage of such an approach lies in having to compare only the
performance hit felt by a given algorithm and/or architecture when the
instance complexity is increased. Furthermore, we propose a methodology
that might not directly translate into the detection of quantum speedup,
but might elucidate whether quantum annealing has a ``quantum
advantage'' over corresponding classical algorithms like simulated
annealing. Our results on a 496-qubit D-Wave Two quantum annealing
device are compared to recently used state-of-the-art thermal simulated
annealing codes.

\end{abstract}

\pacs{75.50.Lk, 75.40.Mg, 05.50.+q, 03.67.Lx}

\maketitle

\section{Introduction}

Optimization plays an integral role across disciplines. Not only does
modern manufacturing and transport heavily depend on efficient
optimization methods to reduce cost and emissions, many fields of
research depend on a multitude of optimization techniques to solve a
wide variety of problems. Similarly, the ever-increasing amount of data
available to mankind means an urgent need for more efficient approaches
in querying, parsing, and mining data, approaches that often depend on
optimization techniques. Within physics-related disciplines alone,
optimization is needed to solve many difficult problems ranging from
frustrated spin systems \cite{hartmann:01,juenger:01,hartmann:04} to
novel approaches in material discovery, as well as the efficient parsing
of high-energy event data or astrophysical spectra. As such, the search
for more efficient optimization approaches is of great importance.
Because the speedup of current silicon-based computing technologies is
slowly coming to an end mostly due to manufacturing and material
constraints \cite{moore:65}, interest in developing faster optimization
methods has shifted to the development of new state-of-the-art
algorithms, as well as novel computing paradigms, e.g., based on quantum
architectures.

Quantum computing \cite{nielsen:00,nishimori:01} and, in particular,
adiabatic quantum optimization
\cite{finnila:94,kadowaki:98,brooke:99,farhi:00,roland:02,santoro:02,das:05,santoro:06,lidar:08,das:08,morita:08,mukherjee:15}
has gained increased momentum since D-Wave Systems Inc.~introduced the
D-Wave Two (DW2) quantum annealing device \cite{johnson:11}. Inspired by
the work of Santoro {\em et al.}~\cite{santoro:02}, multiple teams have
attempted to demonstrate that quantum adiabatic optimization---or
quantum annealing (QA) \cite{amin:09,young:10,hen:11,matsuda:09}---has
advantages over conventional thermal optimization techniques, such as,
for example, simulated annealing (SA) \cite{kirkpatrick:83}.  The idea
behind QA is to adiabatically quench quantum fluctuations to optimize a
cost function (Hamiltonian) of a given complex optimization problem.
Potentially, the wave function of the problem might be able to quantum
tunnel through barriers in the free-energy landscape, i.e., QA might be
able to outperform other approaches like SA where temperature
fluctuations are slowly reduced to find the optimum. Towards the end of
the annealing schedule in SA, when these temperature fluctuations are
small, the system is unable to overcome free-energy barriers and,
especially for problems with rough energy landscapes such as in spin
glasses \cite{binder:86,stein:13} and related problems, it might become
trapped in metastable states, thus missing the true optimum of the
problem.

The fact that a broad range of hallmark optimization problems, such as
the satisfiability problem ($k$-SAT), the number partitioning problem,
vertex covers, knapsack problems, coloring problems, the traveling
salesman problem, etc.~can be mapped onto quadratic unconstrained binary
optimization problems \cite{lucas:14}, means that devices that are
tailored to solve these, such as the DW2, could revolutionize today's
optimization efforts. Although not a fully programmable universal
quantum computer, the D-Wave device represents a sizable advance in
(quantum) computing.

The seminal work of R{\o}nnow {\em et al}.~\cite{ronnow:14a} took great
care and detail in defining the notion of {\em quantum speedup}. While
at the moment the demonstration of {\em strong quantum speedup} remains
a distant goal, the detection of {\em limited quantum speedup}
\cite{comment:limited}---a speedup relative to a given corresponding
classical algorithm such as SA---seems more graspable. The number of
studies (see, for example,
Refs.~\cite{nagaj:12,pudenz:13,boixo:13a,boixo:14,boixo:14,ronnow:14a})
attempting to detect quantum speedup is growing at a fast pace;
however, the definite detection of quantum speedup remains elusive. So
why, despite these large efforts, does quantum speedup remain to be
demonstrated?  Potentially, there are many reasons why this might be the
case. On one hand the complex circuitry, combined with the extreme
fragility of quantum states to perturbations might be a source of
decoherence and thus loss of any advantage over conventional techniques.
On the other hand, the systems currently available (maximally $512$
qubits on DW2, soon up to $\sim 1000$) might be too small for the
benchmarks to be in the asymptotic scaling regime. However, a more
mundane reason that is relatively easy to fix is the choice of the wrong
benchmark problem. In Ref.~\cite{katzgraber:14}, Katzgraber {\em et
al}.~demonstrated that the native benchmark to search for quantum
speedup on a device like the DW2---an Ising spin glass with discrete
uncorrelated disorder---is likely a problem that not only might be too
easy to detect any speedup (think of two world-class skiers on a bunny
slope), but the energy landscape of a spin glass on the DW2 Chimera
topology \cite{bunyk:14} might actually favor thermal approaches like
SA, simply because the spin-glass state exists only at zero temperature.
Furthermore, the use of either bimodal or uniform range-$k$ disorder
\cite{pudenz:13,boixo:13a,boixo:14,boixo:14,ronnow:14a} creates an
energy landscape that has a {\em huge} number of configurations that
minimize the cost function. As such, any method like SA run with
multiple restarts will naturally excel in optimizing such a problem.
Attempts to mitigate this issue by {\em planting} solutions
\cite{hen:15} delivers problem instances that might not be challenging
enough for both classical algorithms and quantum devices alike.

To overcome the limitations imposed by the small size of current
devices, it is imperative to use a {\em native} benchmark problem that
uses as many qubits $N$ as possible on the device. Any {\em embedding}
of a potentially harder problem \cite{venturelli:15} will further reduce
the number of logical qubits, thus pushing the asymptotic regime farther
away. Furthermore, it is hard to mitigate the effects of noise on both
qubits and couplers without improving manufacturing.  However, it is
considerably easier to design hard benchmark instances that attempt to
work around the flaws and limitations of the DW2 architecture.
Reference \cite{zhu:15a} focuses on designing instance problems that are
affected as little as possible by the chip's intrinsic noise. Here, we
present a simple road map that uses insights from the study of spin
glasses to design hard, as well as tunable, benchmark instances.

In addition, we propose to search for quantum advantages over classical
architectures not only by comparing to state-of-the-art classical
algorithms \cite{isakov:15}, but by studying the effects of tuning the
instance complexity for a given type of disorder on both classical and
quantum approaches. By studying the performance hit felt by the
different approaches on carefully tailored problems with a free-energy
landscape that is either dominated by large barriers or is reminiscent
of a ferromagnetic system, further insights into the nature of quantum
annealing devices can be gained.  To perform a fair comparison across
instances, here we fix the ground-state degeneracy (ideally) to $1$ (or
as low as possible) and vary the complexity of the free-energy landscape
by using the spin-glass order parameter distribution as a proxy to the
dominant features of the landscape \cite{yucesoy:13,comment:pq}. We show
that, indeed, the spin-glass order parameter distribution produces
tunable instances, and that predictions from the study of spin glasses
on the complexity of the energy landscape allows us to produce problems
on average considerably harder than any previous study.

We emphasize that we are {\em not} attempting to perform a scaling
analysis as done in previous studies, simply because we believe that the
currently accessible system sizes of up to $512$ qubits are too small to
be in the asymptotic limit \cite{comment:fss}. We base this statement on
previous simulations of two-dimensional Ising spin glasses on a square
lattice at zero temperature with discrete disorder \cite{campbell:04}
where corrections to scaling due to the finite system sizes were very
strong for systems with $\sim 10^3$ spins.

Our results show that the DW2 device is outperformed at finding the
ground state by classical state-of-the-art optimization
algorithms. However, there is a potential signature that the DW2
device might be able to optimize certain classes of carefully designed
native spin-glass problems more efficiently than the classical
counterpart SA, especially if noise is reduced. This suggests that the
DW2 device potentially has a ``quantum advantage'' over corresponding
classical algorithms like SA for certain problems.  In addition, there
are signs that the DW2 device might in some cases be more effective at
generating low-lying states, as opposed to strict ground states than
SA. Finally, our results suggest that ``classical computational
hardness'' in spin glasses seems to carry over to quantum annealing
devices, therefore facilitating the design of spin-glass-based
instances.  The day that quantum annealing machines have lower
noise levels, higher connectivity to enable the simple embedding of
spin-glass problems with, e.g., a finite transition temperature
\cite{katzgraber:14,venturelli:15}, or a larger numbers of qubits, a
combination of the approach presented in Ref.~\cite{ronnow:14a}, with
error-correction techniques \cite{pudenz:13,pudenz:15}, and designer
instances described in this work will likely show if quantum speedup
is myth or reality.

The paper is structured as follows. In Sec.~\ref{sec:native}, we
introduce the native benchmark problem, followed by a detailed
description of the limitations of current approaches as well as how we
design hard instance problems in Sec.~\ref{sec:selection}. Section
\ref{sec:results} summarizes results on both the DW2 device, as well as
classical simulation codes, followed by a discussion and summary.
Appendix \ref{app:dw2} outlines our experimental methodology on the DW2
device housed at D-Wave Systems Inc., followed by simulation details in
Appendix \ref{app:ptmc} and numerical results in Appendix
\ref{app:results}.  Appendix \ref{app:other} summarizes less fruitful
efforts experimenting with other instance classes.

\section{Native Benchmark: Spin Glasses}
\label{sec:native}

We illustrate our benchmarking ideas using the D-Wave Systems,
Inc.,~D-Wave Two quantum annealing machine \cite{comment:d-wave}. The
{\em native} benchmark problem for the DW2 device is an Ising spin glass
\cite{binder:86,nishimori:01,stein:13,lucas:14} defined on the Chimera
topology of the system \cite{bunyk:14},
\begin{equation}
{\mathcal H} =  -\sum_{\{i,j\} \in {\mathcal V}}J_{ij}S^z_i S^z_j 
                - \sum_{i \in {\mathcal V}} S^z_i h_i \, .
\label{eq:ham}
\end{equation}
The $N$ Ising spins $S^z_i \in \{\pm 1\}$ are defined on the vertices
${\mathcal V}$ of the Chimera lattice (see Fig.~\ref{fig:chimera}) and
can be coupled to a (local) field $h_i$.  The sum is over all edges
${\mathcal E}$ connecting vertices $\{i,j\} \in {\mathcal V}$.  In this
study we set $h_i = 0$ $\forall i$.

We emphasize that it is of paramount importance to study {\em native}
problems that use as many qubits as possible to prevent overhead that
might yield smaller embedded problems. At the moment, with approximately
$500$ (soon $1000$) qubits at hand, it will be difficult to detect any
quantum speedup. As such, our focus does not lie in performing a
detailed scaling analysis with the problem size $N$, but to show how to
select tunable hard problems that have the {\em same} disorder
distribution, i.e., have the same strengths or weaknesses with respect
to the intrinsic noise found in these devices. Tuning the complexity of
the problem instances will then allow for a systematic testing of any
potential advantages or disadvantages that the DW2 device might have
over other architectures and/or simulation approaches.  Note that in
this study we disregard the effects of noise on the couplers and qubits
and will report on these in a subsequent publication with strategies on
how to mitigate the effects of perturbed problem Hamiltonians
\cite{zhu:15a}. However, for the generated problems, the resilience to
noise (robustness to perturbations) on the qubits and couplers is
roughly similar and mostly agrees within error bars for the different
instance subclasses that use interactions based on Sidon sets
\cite{sidon:32}; see Sec.~\ref{sec:tune} for details.  This means that
the noise of the DW2 does not affect our results.

\section{Designing Hard Instances}
\label{sec:selection}

We start by describing the shortcomings of previous instances to detect
quantum speedup and then outline our approach to produce tunable, hard
instances.

In Ref.~\cite{katzgraber:14} it was shown that a spin glass on the
Chimera topology has a zero-temperature phase transition. Although the
worst case complexity of finding a ground state of an Ising spin glass
on the Chimera graph falls into the NP hard class, performing any
minimization of the energy based on any annealing approach will likely
have a rather simple phase space to traverse for small system sizes
because dominant barriers will not be as pronounced. Embedding problems
that have a finite-temperature spin-glass transition is difficult,
mainly due to the large overhead; i.e., only systems with few logical
qubits can be studied because many physical qubits are needed to emulate
long-range interactions. Because the resulting systems are small, the
problems are far from the asymptotic regime to detect any quantum
speedup in a scaling analysis.

A more promising route is thus to use insights from the study of spin
glasses and carefully design the interactions between the qubits on the
native Chimera graph, such that the problems are as hard as possible in
order to challenge any optimization approach.

\subsection{Problems with current approaches}
\label{sec:prob}

In addition to a restrictive geometry, the D-Wave hardware has clear
restrictions as to what values the interactions between the spins can
have. This is rather limiting and, as such, only discrete and
well-separated values of the couplers can be set. The simplest approach
used in previous studies
\cite{pudenz:13,boixo:13a,boixo:14,boixo:14,ronnow:14a} is to select the
disorder from a bimodal distribution, i.e., $J_{ij} \in \{\pm 1\}$ (we
shall refer to these as U$_1$), followed by uniform range-$k$ problems
where the interactions $J_{ij}$ are chosen from the integer set $\{\pm
1, \pm2, \ldots, \pm k\}$. We refer to the latter as U$_k$. The problem
with these choices for systems up to $N = 512$ variables is the huge
degeneracy of the ground states that yields again benchmarks too simple
to challenge any optimization approach (see Sec.~\ref{sec:results}).  A
simple analogy to this problem is a game of golf where the green has,
for example, $10^7$ holes. Hitting a hole in one is a trivial task!
However, having a course with only one hole makes the sport truly
challenging. As such, we design herein problems that---within the
hardware restrictions of the machine---have a unique configuration that
minimizes the Hamiltonian in Eq.~\eqref{eq:ham}.

Other approaches \cite{neuhaus:14,hen:15} using planted solutions suffer
from similar problems: While the instances are harder than for the
problems in the U$_k$ class, they often still have a large degeneracy
and their complexity is not high enough for the current available
systems of up to $\sim 10^3$ qubits. In particular, the very careful
work presented in Ref.~\cite{hen:15} shows a clear easy-hard-easy
transition of the planted $k$-SAT solutions that could be exploited to
generate hard instances. However, one problem that these instances have
is that the disorder is not drawn from a particular distribution; i.e.,
two different planted $k$-SAT instances will likely have a very
different (classical) energy spectrum and thus also be differently
susceptible to the intrinsic noise found in the DW2 device
\cite{comment:king}. Furthermore, we perform experiments with
planted $k$-SAT solutions as presented in Ref.~\cite{hen:15} using the
benchmark codes in Ref.~\cite{isakov:15} and find that these instances
are at times easier than the ones in the U$_1$ class. The authors of
Ref.~\cite{hen:15} do emphasize that harder problems must be designed to
allow for the optimization of the annealing time, as well as the need to
find problems where the benefits of quantum annealing can be assessed
{\em ahead of time}.

Finally, setting the spin-spin interactions within the K$_{4,4}$ unit
cell of Chimera (see Fig.~\ref{fig:chimera}) to be of larger magnitude
than those between the cells (often referred to as ``cluster problems'')
has given DW2 an advantage over classical codes in a scaling analysis
\cite{boixo:14a} when cluster Monte Carlo updates are not allowed.
However, by design, simulated annealing (and any other Monte Carlo-like
simple-sampling variation) will have a large disadvantage. The addition
of simple clusterlike moves would again give classical approaches the
upper hand and, as such, these approaches are not a viable route to
detect any speedup, especially because they are unphysical.

\subsection{Designing tunable hard instances}
\label{sec:tune}

Our approach to generate hard instances capitalizes on the similarity
between classical hardness of spin-glass-like problems and quantum
hardness. In Fig.~6 of Ref.~\cite{yucesoy:13}, it was shown in detail how
the ``mixing''  or ``autocorrelation'' time strongly correlates to the
complexity of the spin-glass order parameter distribution while
performing the simulations with state-of-the-art parallel tempering
Monte Carlo methods \cite{hukushima:96,geyer:91,katzgraber:06a}.
Autocorrelation times uniquely characterize the time a classical
algorithm needs to completely decorrelate the system. As such, the time
can be used as an indirect proxy of the time complexity of a particular
disorder instance.

In spin glasses, order is measured by comparing two copies of the system
with the same disorder \cite{binder:86}. For simplicity, we set
$S_i^z \equiv S_i$, because we are studying the system classically. In
that case, the overlap between two replicas $\alpha$ and $\beta$ with
the same disorder ${\mathcal J}$ but independent Markov chains is
defined via
\begin{equation}
q = \frac{1}{N}\sum_{i = 1}^N S_i^\alpha S_i^\beta ,
\label{eq:q}
\end{equation}
where the sum is over all spins $N$. One can then study the distribution
of the order parameter $P(q)$ which characterizes a given disorder
instance ${\mathcal J}$. After a disorder average $[\cdots]_{\rm av}$
over many instances ${\mathcal P}(q) = [P(q)]_{\rm av}$ displays a
single peak around $q \sim 0$ for high temperatures. For $T \to 0$ two
peaks at $\pm q_{\rm EA}$ emerge \cite{edwards:75,parisi:83}, a
characteristic signature of a broken symmetry. However, for a given
instance the structure of the distribution $P(q)$ can be rather complex
and can have multiple peaks at different values of $q$ in addition to
the two dominant peaks at $\pm q_{\rm EA}$. Individual peaks can be
identified with pairs of dominant valleys in the (free-) energy landscape
\cite{stein:13}. When these peaks are close to $q \approx 0$, one can
assume that a thick barrier separates these valleys, whereas when the
peaks are close the barriers are typically thin.

\begin{figure}[h]
\begin{center}
\includegraphics[width=1.00\columnwidth]{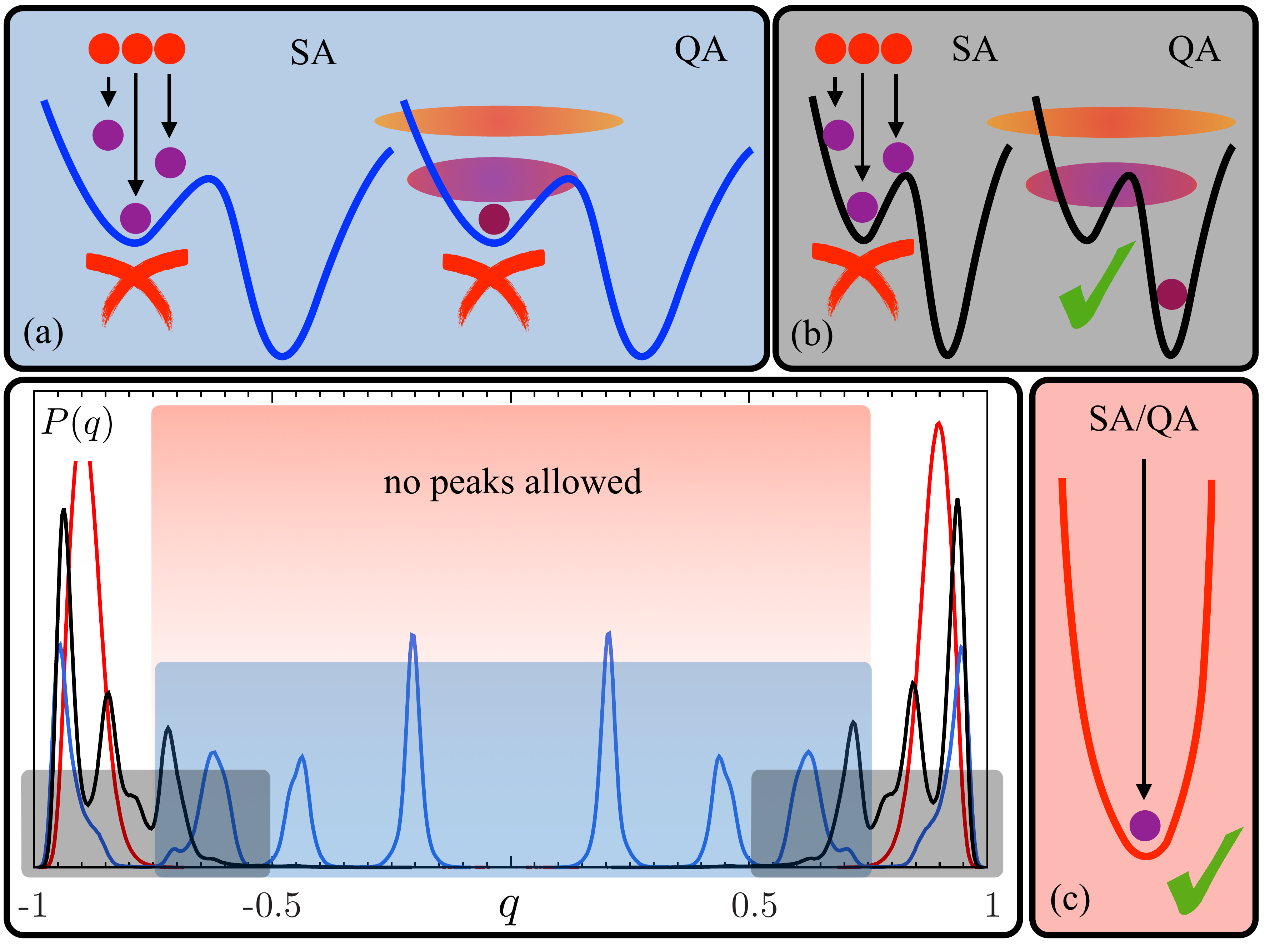}
\end{center}
\caption{
Main panel: Overlap distribution $P(q)$ for three characteristic
disorder instances. For the {\em hard instances with thick barriers}, we
choose instances in the central [blue (dark)] box that have features that
extend outside this domain. Based on classical simulations, we expect
these instances to be on average among the hardest. In (a), we show
the expected outcome of experiments with both simulated annealing (SA)
and quantum annealing on the DW2 device (QA). Because the barriers are
large and thick, we expect both classical and quantum approaches to have
difficulties. In (b), we illustrate the expected behavior when the
barriers are thin, i.e., double peaks (or more) that protrude from the
dark boxes in the region $|q|> 0.5$. The features in the energy
landscape of these {\em hard instances with thin barriers} are still
very pronounced, but we expect the barriers to be thinner than in 
(a). While SA should show little to no advantage when the barriers
remain high but are thinner, if the DW2 device has any quantum
advantage, it might be able to overcome these barriers. Finally, we
study instances that have no features for $|q| < 0.75$ (large red box in
the main panel) and only have a single peak at $\pm q_{\rm EA}$. These
{\em (hard) instances with small barriers} have the simplest energy
landscape (c) with mostly only one dominant feature. As such, we
expect any annealing approach to efficiently find the optimum of the
problem (on average). Note that these are cartoons intended to illustrate
the different instance classes and do not represent actual data.
}
\label{fig:pq}
\end{figure}

Reference \cite{yucesoy:13} showed that when the distribution $P(q)$ has
large support for an area close to $q = 0$, then the autocorrelation
times were typically larger than when the support around $q = 0$ is
close to zero. As such, by measuring the distribution function $P(q)$,
we can predict approximately the time complexity of a particular
disorder instance \cite{comment:pq}. This is illustrated in the main
panel (bottom left) of Fig.~\ref{fig:pq}. There, three characteristic
instances are shown (color coded). An instance with many peaks close to
$q = 0$ will typically be computationally harder than one that has only
two peaks at $q \sim 1$ (red line). Our experiments (shown herein) on the
DW2 device show that, indeed, the complexity of an instance can be tuned
by studying the structure of $P(q)$ where the distance between two
dominant peaks corresponds roughly to the barrier thickness in phase
space and the relative depth between the peaks and maxima can be
interpreted approximately as the barrier depth. While we are confident
that there is a clear correlation between the distance $\Delta q$ of two
well-defined peaks and the thickness of barriers in the energy
landscape, the correlation of the depth between the peaks and the height
of the barriers remains to be tested experimentally by a more precise
mining of the data. However, if the depth between the peaks is nonzero,
then it is safe to assume that there is some relatively trivial path
that connects the valleys \cite{comment:paths}.

In addition to selecting instances according to the complexity of the
phase space by studying the behavior of the spin-glass order parameter
distribution, we estimate the number of configurations for a given
instance that minimize the Hamiltonian in Eq.~\eqref{eq:ham}. The goal
is to make the problem as difficult as possible by restricting the
number of minimizing configurations ideally to one, i.e., a  unique
ground state. To estimate the number of ground-state configurations a
given instance has, we use the method pioneered in
Refs.~\cite{moreno:03,katzgraber:03f} where states at very low
temperatures are sampled with parallel tempering Monte Carlo techniques.
Once the ground-state energy is found, a histogram with minimizing
configurations is created (indexed by translating the binary
configuration string to a number) and sampled until every bin has at
least $50$ hits. We make sure that we find the true ground-state energy
by studying every instance with different simulational heuristics.
However, we cannot be completely certain that we have found {\em all}
configurations that minimize the Hamiltonian, simply because in some
cases this number can be huge (in the worst case $2^N$). Having exactly
one ground state is not a necessary condition to generate a hard
problem. However, if our efficient low-temperature search is unable to
find more states that minimize the cost function, it will be unlikely
that other methods will.

A large source of degeneracy in an Ising Hamiltonian is due to zero
local fields. The Hamiltonian in Eq.~\eqref{eq:ham} can be written as a
single-spin expression, namely,
\begin{equation}
{\mathcal H} = \sum_{i \in {\mathcal V}}^N {\mathcal F}_i S_i ,
\label{eq:hamsi}
\end{equation}
where the local fields ${\mathcal F}_i$ are given by
\begin{equation}
{\mathcal F}_i = - \sum_{j\neq i} J_{ij}S_j - h_i .
\label{eq:lf}
\end{equation}
Whenever for a given disorder ${\mathcal F}_i = 0$, spin $S_i$ can take
any value without influencing the energy of the system. Therefore, if a
given disorder instance has $k$ spins where ${\mathcal F}_i = 0$, the
degeneracy of the ground state will grow by a factor $2^k$.  To prevent
this from happening, we need to choose the disorder from a distribution
that---within the restrictions of the device---minimizes the cases where
the local fields are zero. The most convenient choice is thus to select
the values of $|J_{ij}|$ from a Sidon set \cite{sidon:32}. In a Sidon
set, the sum of two members of the set gives a number that is not part
of the set. For example, the set $\{2,5,10\}$ is a Sidon set because the
pairwise sum of members of the set never adds up to a member of the set.
This is not the case  for $\{2,5,7\}$, where $2+5 = 7$.

To illustrate our ideas, we choose the interactions between the spins
from the Sidon set S$_{28}$
\begin{equation}
J_{ij} \in \{\pm 8/28,\pm 13/28, \pm 19/28,  \pm 28/28\} ,
\label{eq:s28}
\end{equation}
where we normalize the interactions to be restricted between $\pm
1$ \cite{comment:s7}.  To select instances with particular properties, we
can therefore generate large numbers of random problems using different
disorder distributions and then mine the data. We first fix the number
of ground-state configurations to $1$, and then we divide the instances
into subclasses by studying the (normalized) overlap distribution $P(q)$
for each instance. For example, we define the following classes:

\begin{itemize}

\item[(a)]{{\em Hard instances with thick barriers}: These are instances
where $P(q) > 5$ for $|q| \le 0.75$. See Fig.~\ref{fig:pq}, main panel. We
are interested in instances that have dominant peaks in the central
(blue/dark) window. Based on classical simulations, we expect these
instances to be on average among the hardest. In particular, we expect
that both simulated, as well as quantum annealing will have trouble
finding the optimum -- see Fig.~\ref{fig:pq}(a).}

\item[(b)]{{\em Hard instances with thin barriers}: These are instances
where $P(q) \approx 0$ for $|q| \le 0.50$ and where $P(q) > 2.5$ for
$|q| \ge 0.5$ with at least two peaks in the range $|q| \in [0.5,1.0]$.
See Fig.~\ref{fig:pq}, main panel. We are interested in instances that
have dominant peaks that are close to each other in the gray boxes close
to $|q| > 0.5$. Based on classical simulations, we expect these
instances to be on average hard, however, not as hard as the instances
with a thick barrier. We expect that while simulated annealing will have
similar problems than with the instances with a thick barrier, quantum
annealing might show an enhanced performance {\em if} the device has
some quantum advantage over classical codes -- see
Fig.~\ref{fig:pq}(b).}

\item[(c)]{{\em (Hard) instances with small barriers}: These are
instances where $P(q) < 0.1$ for $|q| \le 0.75$. The overlap distribution
is reminiscent of a ferromagnet at low temperature. In this case no
peaks are allowed in the large central (red/light) box of
Fig.~\ref{fig:pq}, main panel. In these instances we expect one dominant
energy valley (up to smaller wiggles), i.e., these should be the easiest
instances on average for any annealing approach. See
Fig.~\ref{fig:pq}(c).}

\end{itemize}

\noindent Note that the individual windows we use are tuned such that
from $10^5$ randomly simulated instances approximately $5000$ match the
aforementioned criteria. After filtering the instances that have more
than one minimizing configuration, we obtain approximately $2500$
instances to experiment with. The detailed simulation strategy, as well
as simulation parameters, are listed in Appendix \ref{app:ptmc}.

Noise on the DW2 device is approximately $5\%$ of a particular external
field (qubit noise) $h$ and $3.5\%$ of a spin-spin interaction (coupler)
$J_{ij}$. For the instances in S$_{28}$, the smallest classical energy
gap is $\Delta E = 2/28$, i.e., slightly larger than the noise found on
the DW2 device.  While this will affect the success probabilities, it
will affect {\em all} instances, either easy or hard, approximately the
same way. To verify this, we perform detailed simulations where we
compute the ground-state energy and configuration of a given instance
with no degeneracy, perturb the couplers and qubits with Gaussian random
noise of a typical strength found in the current DW2 device, and
recompute the ground-state configuration. We apply $10$ noise gauges and
compute how stable the different instance subclasses defined below are
on average. Our results show that all Sidon-set-based instance
subclasses with different barrier thicknesses are affected {\em
similarly} by the intrinsic noise of the device (not shown). As such,
when comparing instance classes, on average a fair comparison is
performed.

\section{Results}
\label{sec:results}

A detailed list of the average success probabilities is given in
Appendix \ref{app:results}.  To make sure that an approximately fair
comparison with a known baseline study is performed, we tune the number
of sweeps for the SA codes \cite{isakov:15} such that the average
success probabilities for SA and the DW2 device are approximately the
same for bimodal disorder.  This is the case for $N_{\rm sw} = 900$
sweeps. Note also that below we quote mainly average success
probabilities. The reason is that for the hardest instance classes the
DW2 device is often unable to minimize the cost function for the number
of runs performed; i.e., a median would be zero and thus deliver no
useful information.  Because probabilities are restricted to be in the
interval $[0,1]$, an average is well defined.

\subsection{The ugly---D-Wave Two fails often}

\begin{figure}[h]
\begin{center}
\includegraphics[width=1.00\columnwidth]{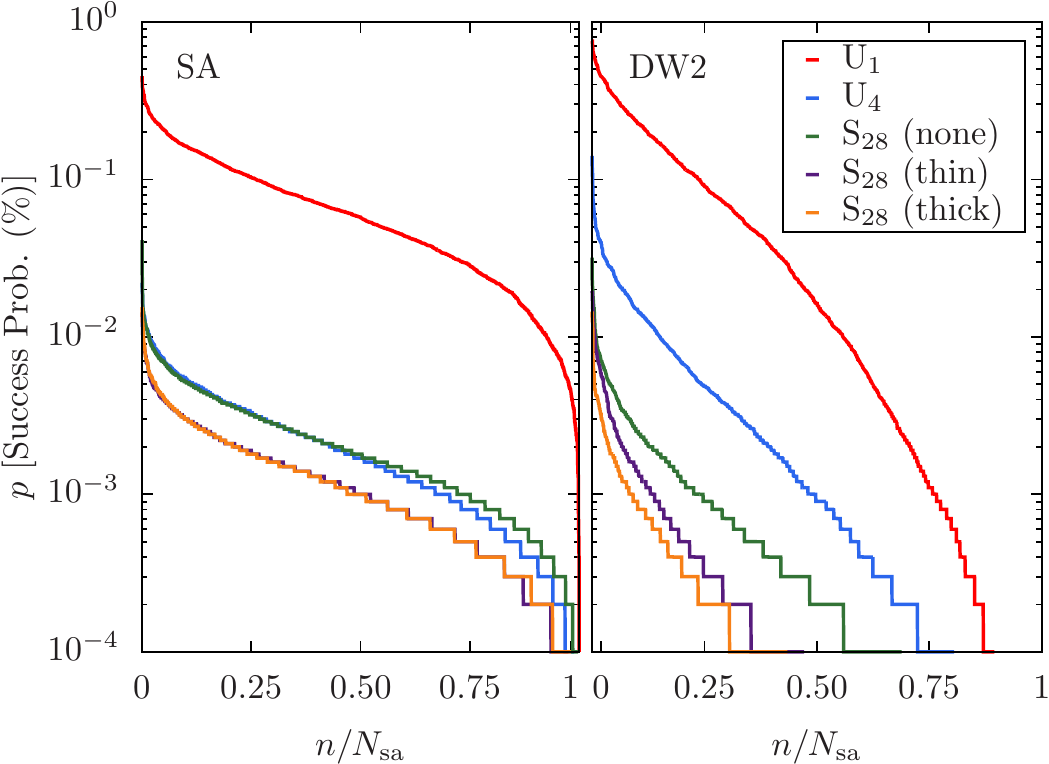}
\end{center}
\caption{
Sorted success probabilities $p$ (after a gauge average) in percent SA
(left) and the DW2 device (right) and different instance classes. The
instance index $n$ is normalized by the number of instances
$N_{\rm sa}$ per class for better viewing. For both cases, bimodal
disorder (U$_{1}$) is the easiest problem class to solve. Although the
shape of these functions is different, the number of sweeps in SA 
are chosen such that on average the success probabilities for the
U$_{1}$ are similar using SA and the DW2.  Using SA, uniform range-$4$
(U$_{4}$) instances are comparable to Sidon instances S$_{28}$ with
small (``none'') barriers. Furthermore, SA does not distinguish between
S$_{28}$ instances with thin and thick barriers. There is a clear
progression in complexity for the different instance classes on the DW2
device. In particular, while SA can solve almost all instances studied,
this is not the case for the DW2. The median success probability for the
hardest instance classes (S$_{28}$) is zero on the DW2 for the number of
runs performed; i.e., the machine would need many more runs to be able
to find the optimum of hard native problems.  Error bars are omitted 
for better viewing.
}
\label{fig:histo}
\end{figure}

Figure \ref{fig:histo} shows sorted success probabilities $p$ for SA
(left) and the DW2 device (right) and different instance classes
normalized by the number of samples $N_{\rm sa}$ studied.  We compare
classes S$_{28}$ with thick, thin, and small barriers with uniform
range-$4$ (U$_{4}$) instances and bimodal disorder (U$_{1}$) used in
previous studies \cite{ronnow:14a}.  The data for the DW2 device show a
clear progression in complexity and, in particular, that the device is
unable to solve many of the harder problems (success probabilities below
$10^{-4}$). The SA simulations using the codes of Ref.~\cite{isakov:15}
show that bimodal disorder is considerably easier than all other
instance classes. Furthermore, for the number of sweeps used, the
complexity of U$_{4}$ is similar to S$_{28}$ with small (``none'')
barriers. Interestingly, the SA codes do not distinguish between
S$_{28}$ instances with thin and thick barriers.  Note that this is not
the case for the DW2 device.

Furthermore, SA can solve a much wider range of instances, as can be
seen by the distributions dropping to zero only close to $n \to N_{\rm
sa}$. This means that while the typical (median) probability to solve a
problem is finite for the SA codes, for the hardest instance classes the
median is zero for the DW2 device.  A double-peaked success behavior of
the quantum annealer is consistent with what has been reported in
Refs.~\cite{boixo:13a,ronnow:14a}, who present it as evidence of quantum
behavior, although the hypothesis has been subsequently challenged by
studies of quasiclassical models \cite{smolin:13,shin:14}.  Finally, we
emphasize that by optimizing the number of sweeps in the SA codes these
can be tuned to outperform the DW2 device for all disorder classes
studied.

\subsection{The bad---Previous instance classes are too easy}

\begin{figure}[h]
\begin{center}
\includegraphics[width=1.00\columnwidth]{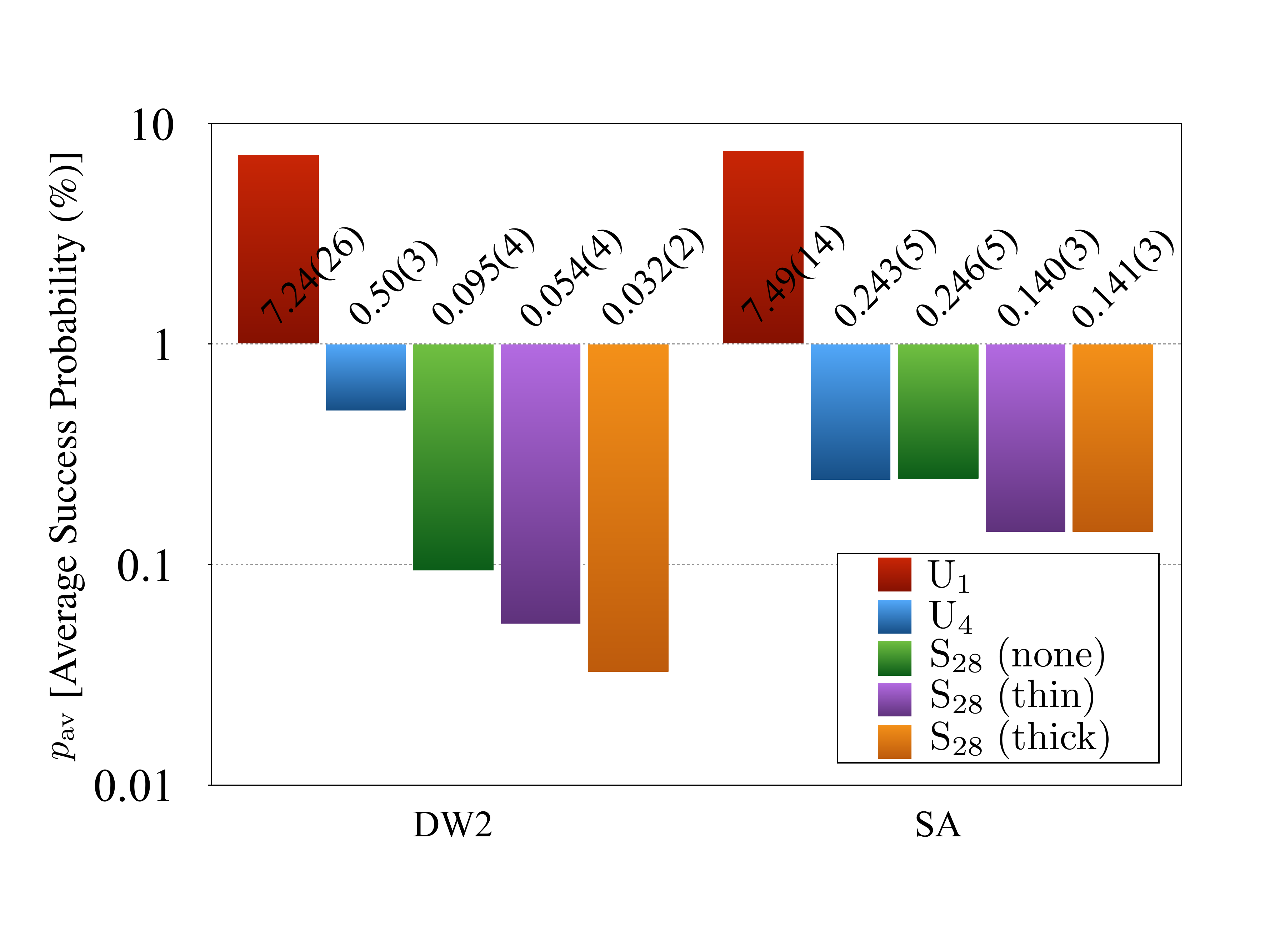}
\end{center}
\vspace*{-2.5em}
\caption{
Average success probabilities $p_{\rm av}$ (after a gauge average) for
DW2 and SA using different types of disorder. In all Sidon instance
classes (S$_{28}$) the classical codes outperform DW2.  Furthermore,
success probabilities for bimodal disorder (U$_{1}$) are much larger
than for any other instance class, therefore suggesting that the
degeneracy produced by bimodal disorder makes this instance class too
easy to detect quantum speedup. Note also that the classical codes, on
average, do not seem to distinguish between instances with thick and
thin barriers. Labels are from left to right.
}
\label{fig:raw}
\end{figure}

\begin{figure}[h]
\begin{center}
\includegraphics[width=1.00\columnwidth]{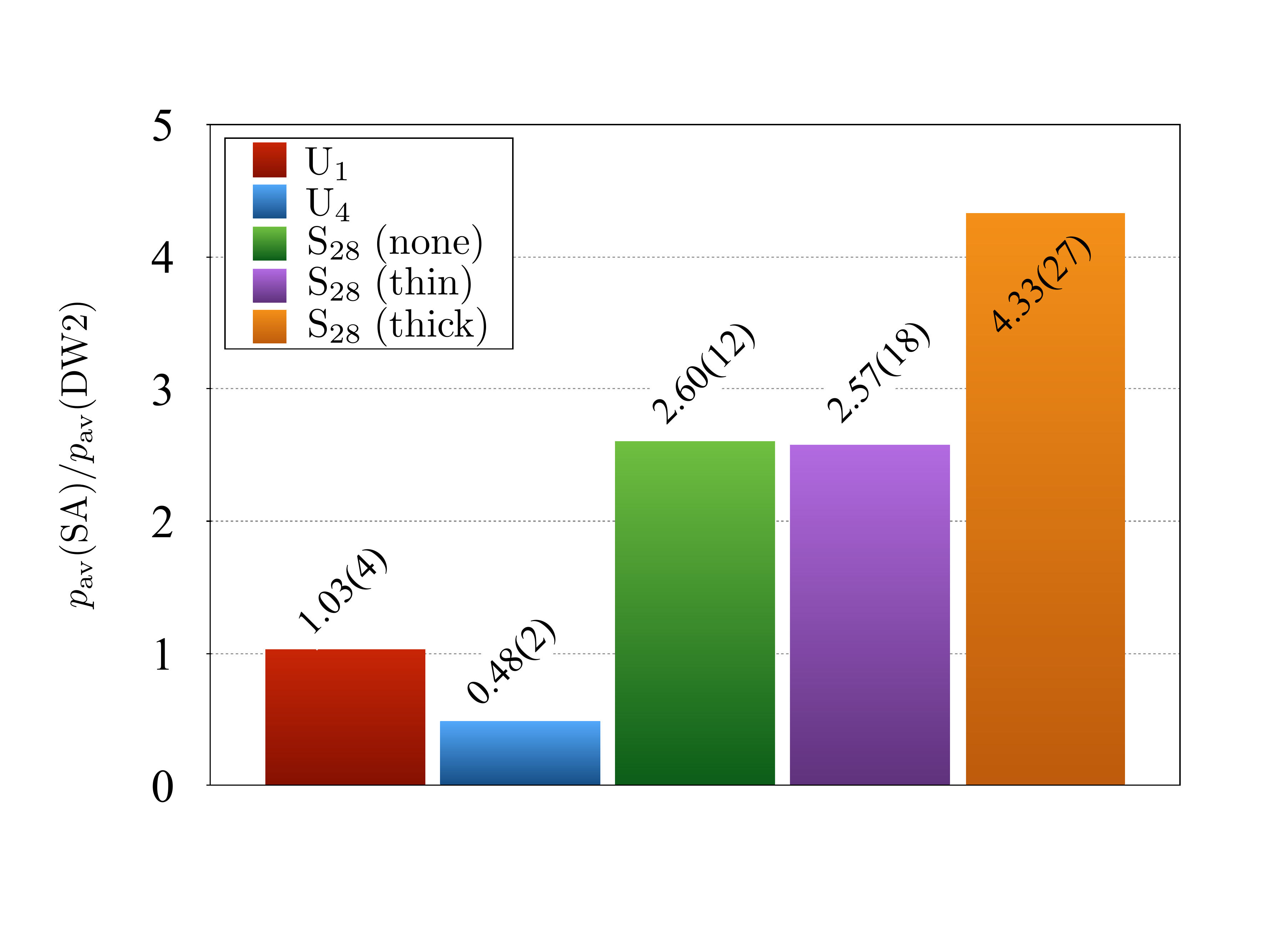}
\end{center}
\vspace*{-2.5em}
\caption{
Ratio between the average success probability for 
SA and DW2, $p_{\rm av}({\rm SA})/p_{\rm av}({\rm DW2})$,
for different disorder classes. In all S$_{28}$ cases, SA outperforms
DW2 when the number of SA sweeps is tuned such that $p_{\rm av}({\rm
SA})/p_{\rm av}({\rm DW2}) \approx 1$ for the bimodal U$_{1}$ class.
Labels are from left to right.
}
\label{fig:ratio}
\end{figure}

Figure \ref{fig:raw} shows averaged (and gauge-averaged) success
probabilities in logarithmic scale for both DW2 and SA for different
instance classes. The data clearly illustrate that the average success
probabilities for bimodal disorder are approximately 1 order of
magnitude larger than any other type of disorder studied.  Note that we
choose the number of sweeps for SA such that the average success
probability in the bimodal class is comparable to the DW2 device. For
the DW2 device, one can clearly see a progression in difficulty between
U$_{1}$, U$_{4}$, as well as the Sidon set S$_{28}$ with small barriers,
followed by the Sidon sets with thin and thick barriers. For the choice
of sweeps in SA, U$_{4}$ is comparable to S$_{28}$ with no dominant
barriers, and the S$_{28}$ instances with thick and thin barriers have
approximately the same average success probabilities.  For all Sidon
instance classes studied, the classical SA simulations outperform DW2
based on raw success probabilities. This is seen in more quantitative
detail in Fig.~\ref{fig:ratio}, which shows the ratio of the average
success probability for SA divided by the average success probability
for DW2 for each instance class.  To establish any quantum speedup, a
system-size scaling is needed. However, the fact that the average
success probabilities for the bimodal disorder for DW2 and the classical
SA codes are much larger than for all other problems suggests that
bimodal disorder (or, more generally, highly degenerate random problems)
is too easy a problem to detect any quantum speedup.  Running any
classical SA code in repetition mode with highly degenerate problems
potentially represents an advantage over any quantum annealing scheme.
Overall, DW2 has far lower average success probabilities on the Sidon
sets. This can be explained by the inherent noise present in the device.
In the Sidon sets the gap to the first excited state is considerably
smaller than for, e.g., bimodal disorder.  As such, solving a
Hamiltonian that is not the target Hamiltonian due to noise-induced
perturbations is likely.  Therefore, in an attempt to filter out these
effects, we study relative probabilities between instance classes and
{\em not} between optimization techniques. Because the problem instances
are randomly generated, one can expect that within a given instance
type, e.g., S$_{28}$, the noise affects all instance classes in a
similar fashion \cite{comment:s7}, as we see in our simulations. This
means also that the difference in the performance of DW2 for S$_{28}$
instances with thick and thin barriers is likely not an artifact of the
chosen values for the couplers.

\subsection{The good---Evidence of a quantum advantage?}

Figure \ref{fig:raw} suggests that---at least with the choice of
annealing parameters made---in the Sidon instance class the classical
codes do not seem to differentiate between thin and thick barriers on
average, whereas DW2 does seem to show an improvement in the average
success probabilities when the barrier thickness is decreased.

\begin{figure}[h]
\begin{center}
\includegraphics[width=1.00\columnwidth]{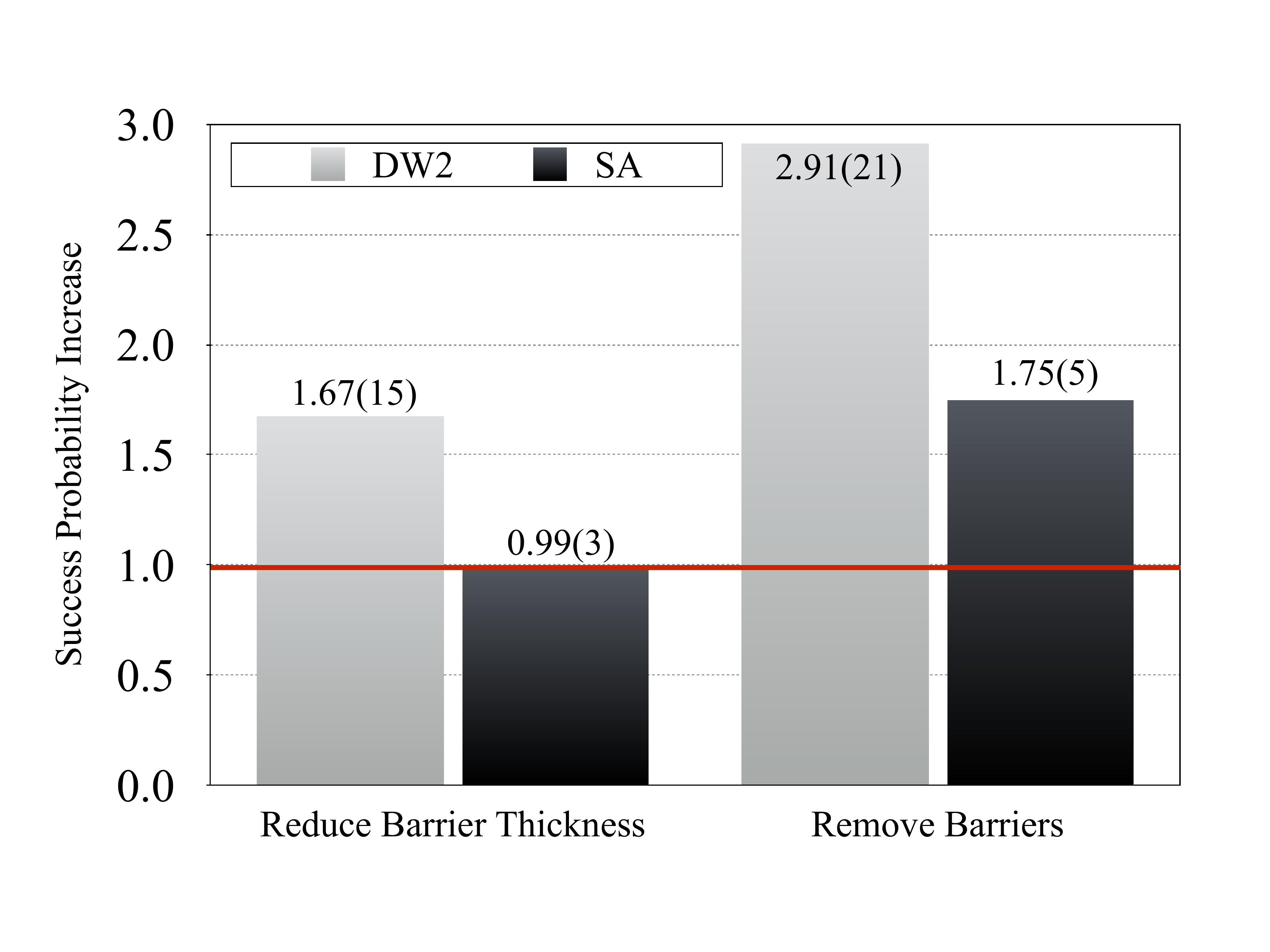}
\end{center}
\vspace*{-2.5em}
\caption{
Average success probability increase when reducing the barrier thickness
(ratio between the average success probabilities for S$_{28}$ thick and
S$_{28}$ thin) and removing the barriers (ratio between the average
success probabilities for S$_{28}$ thick and S$_{28}$ none). While in the
latter case both classical algorithms and the quantum annealer show a
performance boost on average, in the former only the quantum annealer
shows improvement.
}
\label{fig:speedup}
\end{figure}

Given the stochastic nature of the classical algorithms, the thickness
of a barrier should have a much weaker effect on the algorithmic
efficiency than its height. We have selected the instances in such a way
that barriers are predominantly tall. Although we have no exact control
at the moment as to how tall these barriers are, we can expect them to
be on average of similar height for both Sidon sets with thin and thick
barriers. However, by selecting instances with peaks in the overlap
distribution at a given distance from each other, we have good control
over the barrier thickness. Figure \ref{fig:speedup} shows the ratio of
average success probabilities when reducing the barrier thickness (left)
and removing dominant barriers (right) for both SA and DW2. While
reducing the barrier thickness has no effect on average on the classical
algorithms, DW2 experiences a performance increase.  To make sure this
is not an artifact of our choice of simulation parameters, we run
the SA codes with both $N_{\rm sw} = 900$ and $2000$ sweeps obtaining
qualitatively the same results. Furthermore, we find no correlation
between the barrier thickness and the effects noisy couplers and qubits
have on the success probabilities for both instance classes.  When
removing dominant barriers altogether, both classical and quantum
algorithms show a noticeable performance increase.  One can, therefore,
surmise that when the barriers are thin enough (and tall) the DW2 device
might experience a quantum advantage over classical approaches.
However, a far more careful and systematic study must be performed
before strong conclusions can be drawn.

\begin{figure}[h]
\begin{center}
\includegraphics[width=1.00\columnwidth]{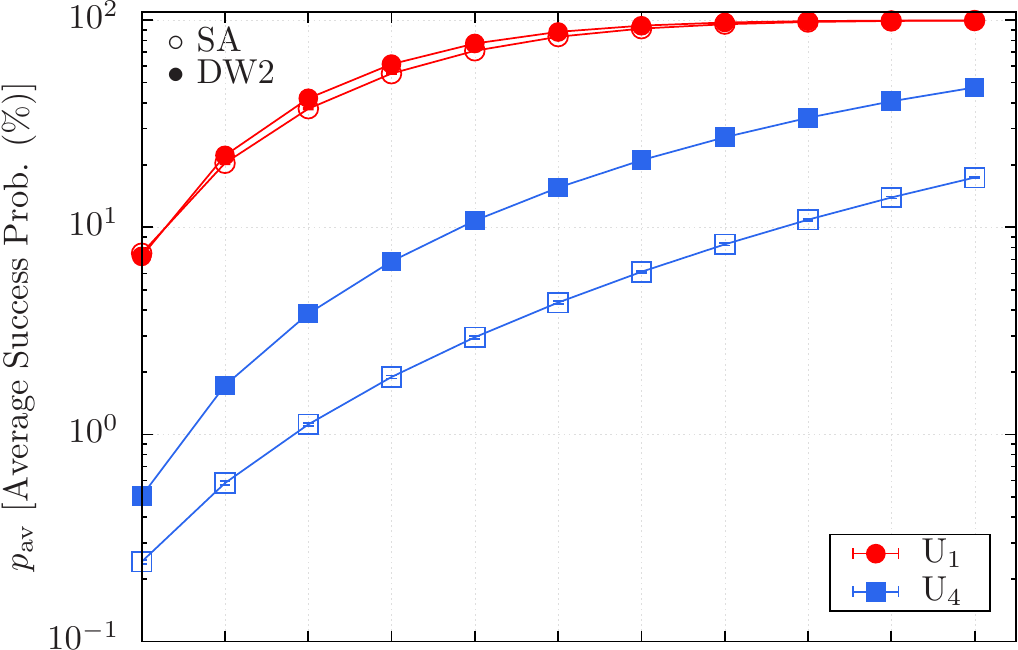}
\includegraphics[width=1.00\columnwidth]{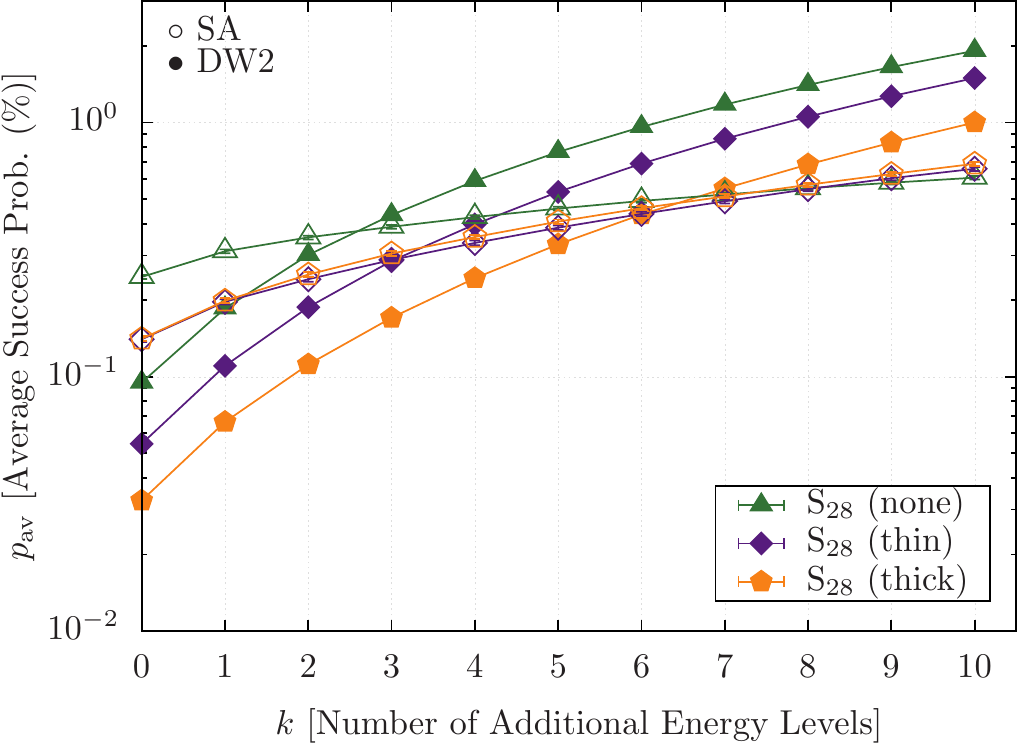}
\end{center}
\caption{
Average success probabilities $p_{\rm av}$ (after a gauge average) for
both DW2 and SA as a function of the number $k$ of low-lying energy
levels above the ground state for different instance classes.  The top
panel shows data for the U$_{1}$ and U$_{4}$ instance classes, whereas
the bottom panel shows data for the S$_{28}$ class. Both panels have the
same horizontal axis; we split up the data for better viewing.  The
trend displayed by DW2 compared to the SA codes for the S$_{28}$ class
suggests that noise might be the dominant source of the overall poor
performance of the DW2 device.
}
\label{fig:relax}
\end{figure}

To gain a deeper understanding of the noise effects that affect the DW2
device, we relax our criterion for a successful optimization run by
allowing the $k$ lowest excited states to count towards a ``successful''
run in the Sidon sets. In this case, the smallest classical energy gap
when flipping a spin is $\Delta E = 2/28 \approx 0.0714$. This should be
compared with the disorder-averaged ground state energy of the system,
i.e, $[E_0]_{\rm av} \approx = -551$. We compute the success
probabilities for energies in the interval $[E_0, E_0 + k \Delta E]$ for
different instance classes using SA and the DW2.  Figure \ref{fig:relax}
shows the average success probabilities as a function of the number of
energy levels $k$.  Although we only fix the average success
probabilities for the U$_1$ class to be similar for DW2 (full symbols)
and SA (empty symbols) and $k = 0$, it seems this result holds for at
least the first $10$ excited states. As can be seen, average success
probabilities increase with an increased inclusion of low-lying energy
levels for all instance classes.  The trend is far more pronounced for
the DW2 device than for SA in the case of the Sidon sets S$_{28}$,
indicating that noise clearly affects the ability of the machine to
detect ground states. Furthermore, note that allowing for the lowest
$10$ energy levels in the S$_{28}$ class corresponds to an increase in
less than 1\% in the overall energy of the system. Averaging over gauges
(i.e., different instances of noise terms in the Hamiltonian) does help
the DW2 device, thus illustrating that an increased performance strongly
depends on reducing noise, and also performing multiple quenches.

Is the DW2 device of any use then? For problems affected by noise due to
device restrictions, the DW2 thus might efficiently deliver low-lying
energy states. This is of particular relevance to problem domains such
as machine learning \cite{bishop:06} and Bayesian statistical analysis
\cite{gelman:03}.

For optimization, the data suggest that error-correction strategies
\cite{pudenz:13} that enhance robustness to noise should be explored in
greater depth. Combined with a hybrid approach that either breaks up the
problem into smaller groups that are easier to tackle
\cite{houdayer:99a,thomas:08,zintchenko:15}, or uses other efficient
computing architectures \cite{belletti:08} to complement the
minimization, the DW2 device (or any other quantum annealing machine)
might be an efficient optimization tool one day.

\section{Discussion}
\label{sec:discussion}

We illustrate that a careful design of the benchmark instances is key
when attempting to detect quantum speedup. In particular, using insights
from the study of spin glasses can help in designing benchmark problems
that are considerably harder than previous attempts, and are tunable.
Noise levels combined with the small number of qubits on the DW2 device
make it difficult to detect any quantum speedup at the moment.  Below,
we attempt to discuss sources of the poor performance of the device as
seen from the spin-glass perspective.

Disordered frustrated binary systems are the {\em native}, likely
hardest, as well as simplest benchmark problems for any new (quantum)
computing paradigm. It is important to consider some of the hallmark
properties of spin glasses that could make it extremely difficult to
detect any (quantum) speedup in the presence of coupler, as well as
local-field qubit noise.

\subsection{Effects of coupler noise} 

The extreme fragility of the spin-glass state was predicted a long
time ago \cite{mckay:82,parisi:84} and analyzed on the basis of scaling
arguments \cite{fisher:86,bray:87}. These scaling arguments predict that
the configurations that dominate the partition function change
drastically and randomly when temperature, local fields, or the
interactions between the spins are modified.  There is strong
(numerical) evidence of disorder chaos (coupler noise) in spin glasses
\cite{kondor:89,neynifle:97,neynifle:98,billoire:00,billoire:02,sasaki:05,katzgraber:07,comment:chicken}.
Therefore, small perturbations of the couplers due to noise might lead
to the destruction of the spin-glass state, as well as to a change of
the problem to be solved. The latter can be alleviated slightly by
performing multiple gauges.  However, the weak chaos regime is dominated
by rare events that can flip large spin domains that can directly
affect experimental results \cite{katzgraber:07}. Increasing the
classical energy gap beyond the noise level of the machine can partially
reduce these effects, however at the cost of producing considerably
easier benchmark instances \cite{zhu:15a}.

One might argue that the minimum classical gap of the Sidon instances
($\Delta E = 2/28$) is too small compared to the machine restrictions
when encoding problems. However, we perform tests with a
different instance class with a larger classical energy gap and where
the couplers are drawn from the Sidon set $\{\pm 5,\pm 6, \pm7\}$,
finding qualitatively similar results.

\subsection{Effects of local-field noise}

In mean-field theory \cite{parisi:80}, an Ising spin-glass system has a
line of transitions in a field \cite{almeida:78}, known as the de
Almeida-Thouless line that separates the paramagnetic phase at high
temperatures and fields from the spin-glass phase at lower temperatures
and fields
\cite{bhatt:85,billoire:03b,barrat:01,takayama:04,houdayer:99,krzakala:01}.
Although the existence of a de Almeida-Thouless line for short-range
spin glasses is still under some debate (see, for example,
Refs.~\cite{leuzzi:09,banos:12,baity:14}), there is vast numerical
evidence for a multitude of geometries and, in particular,
low-dimensional systems  that the spin-glass state is strongly affected
by any longitudinal (random) fields
\cite{young:04,katzgraber:05c,katzgraber:09b,larson:13}. As for the case
of disorder chaos in spin glasses, the spin-glass state can be easily
affected by the intrinsic qubit noise of the DW2 device. Therefore, it
might be plausible that, again, the high levels of noise might reduce
the success probabilities because the studied system is perturbed and
dominant barriers are affected.

\section{Summary and conclusions}
\label{sec:conclusions}

We find that for most disorder types studied, DW2 is systematically
slower at finding the ground state than the state-of-the-art classical
SA codes developed by Isakov {\em et al.}~\cite{isakov:15}. Note that,
by optimizing the number of sweeps in the SA codes, these can be tuned
to outperform the DW2 device for all disorder classes studied.
Although this might be discouraging at first, we argue that an
improved machine calibration \cite{perdomo:15}, noise reduction
\cite{perdomo:15a}, and the ability to likewise optimize the quantum
annealing schedule combined with larger system sizes and tailored
spin-glass problems might help in the quest for quantum speedup.  We
also show that a ``classically computationally hard'' problem seems to
typically also be a hard problem for the quantum annealing device.
However, it could also be that the DW2 device is a thermal annealer
\cite{smolin:13,smith:13,shin:14,lanting:14,albash:15,albash:15a} in
disguise.

For the hardest Sidon instances the DW2 device does show a promising
trend when the success constraints are relaxed. Furthermore, reducing
the thickness between barriers in the free-energy landscapes suggests
that for the large Sidon instances studied some quantum advantage might
be present. However, this would not be enough to deem the hardware to be
efficient, especially because it is unclear if this effect persists for
larger problem sizes. We conclude by stressing that a careful design of
benchmark instances is key to detecting quantum speedup
\cite{ronnow:14a} or any quantum advantage a novel quantum annealing
device might have. We thus expect that a combination of the
methodologies outlined in this work with the approach outlined in
Ref.~\cite{ronnow:14a} that defines the notion of ``quantum speedup'' in
detail, combined with better hardware (and maybe quantum error
correction \cite{pudenz:13,pudenz:15}), will finally show whether or not
quantum annealing has an advantage over classical thermal annealing.

\begin{acknowledgments}

We are grateful to M.~Amin, R.~S.~Andrist, J.~Job, A.~King, D.~Lidar,
J.~Machta, W.~Macready, O.~Melchert, C.~Moore, A.~Perdomo-Ortiz,
T.~F.~R{\o}nnow, M.~Troyer, D.~Venturelli, and I.~Zintchenko for many
discussions. We especially thank I.~Zintchenko for performing
double-blind sanity checks for us with his codes and O.~Melchert for
computing the misfit parameter \cite{kobe:95} for our instances on
Chimera and for performing detailed studies that link the structure of
the spin-overlap distribution directly to the free-energy landscape
\cite{melchert:15}.  H.~G.~K.~acknowledges support from the NSF (Grant
No.~DMR-1151387) and thanks the City University of New York
at Staten Island for an uneventful job interview that culminated with
him attending a very interesting Mathematics Department colloquium where
he learned about Sidon sets.  H.~G.~K.~would also like to thank
E.~Schmeichel for making this paper happen. We thank the Texas Advanced
Computing Center (TACC) at The University of Texas at Austin for
providing HPC resources (Stampede Cluster), ETH Zurich for CPU time on
the Brutus and Euler clusters, and Texas A\&M University for access to
their Ada, Eos and Lonestar clusters. We especially thank O.~Byrde for
beta access to the Euler cluster, as well as S.~Vellas and F.~Dang for
beta access to the Ada cluster. This research is based upon work
supported in part by the Office of the Director of National Intelligence
(ODNI), Intelligence Advanced Research Projects Activity (IARPA), via
MIT Lincoln Laboratory Air Force Contract No.~FA8721-05-C-0002.  The
views and conclusions contained herein are those of the authors and
should not be interpreted as necessarily representing the official
policies or endorsements, either expressed or implied, of ODNI, IARPA,
or the U.S.~Government.  The U.S.~Government is authorized to reproduce
and distribute reprints for Governmental purpose notwithstanding any
copyright annotation thereon.

\end{acknowledgments}

\appendix

\section{D-Wave Two quantum annealer description}
\label{app:dw2}

\begin{figure}[h]
\begin{center}
\includegraphics[width=1.00\columnwidth]{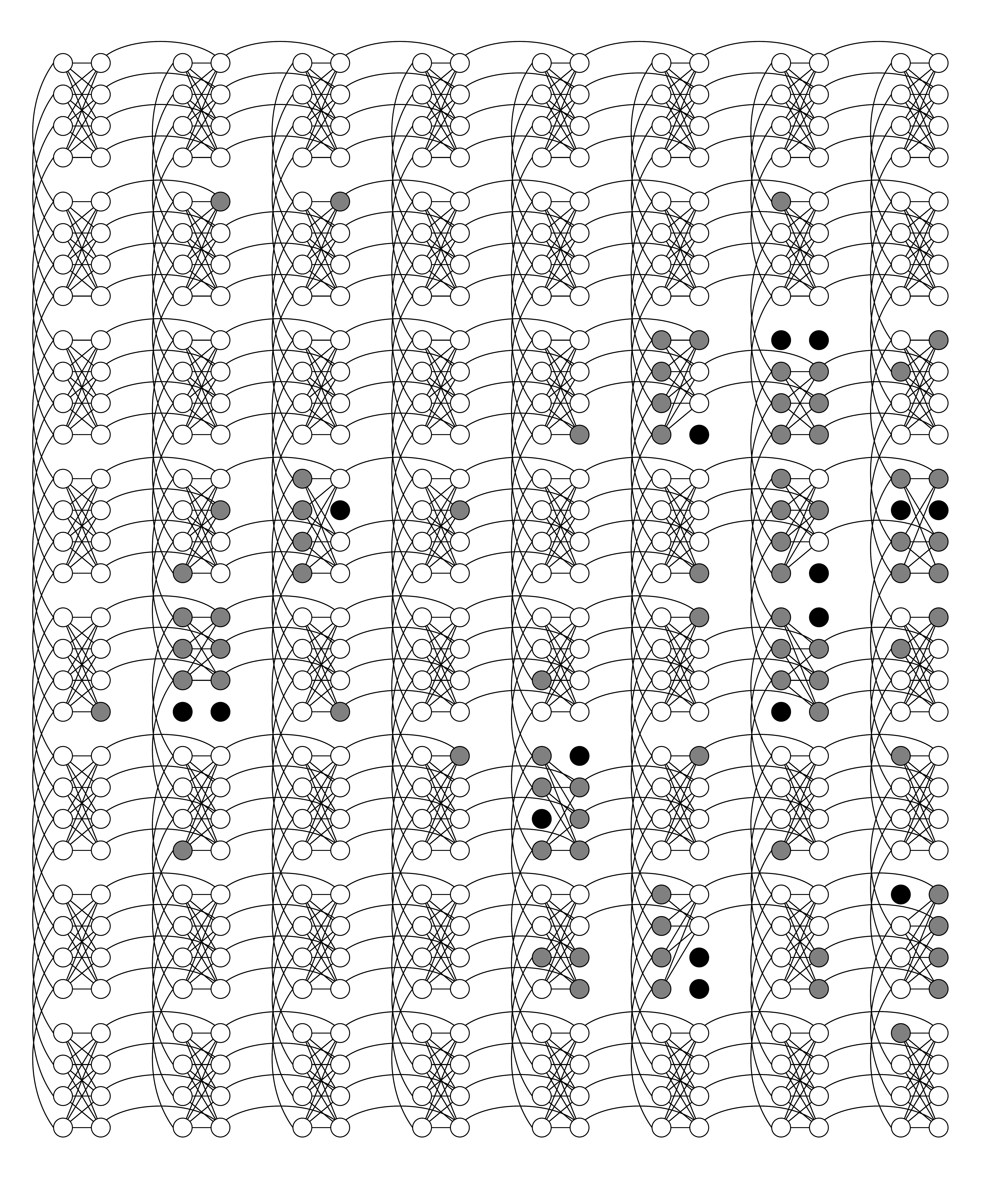}
\end{center}
\caption{
Adjacency matrix of the D-Wave Two chip used in this study. Circles
represent the individual qubits and lines the couplers. White circles
represent fully functional qubits, whereas light gray circles represent
working qubits with missing couplers. Broken qubits are represented by
dark circles ($16$).  This means that the total number of working qubits 
is $496$.
}
\label{fig:chimera}
\end{figure}

\begin{figure}
\centering
\hspace*{-2em}\includegraphics[scale=0.5]{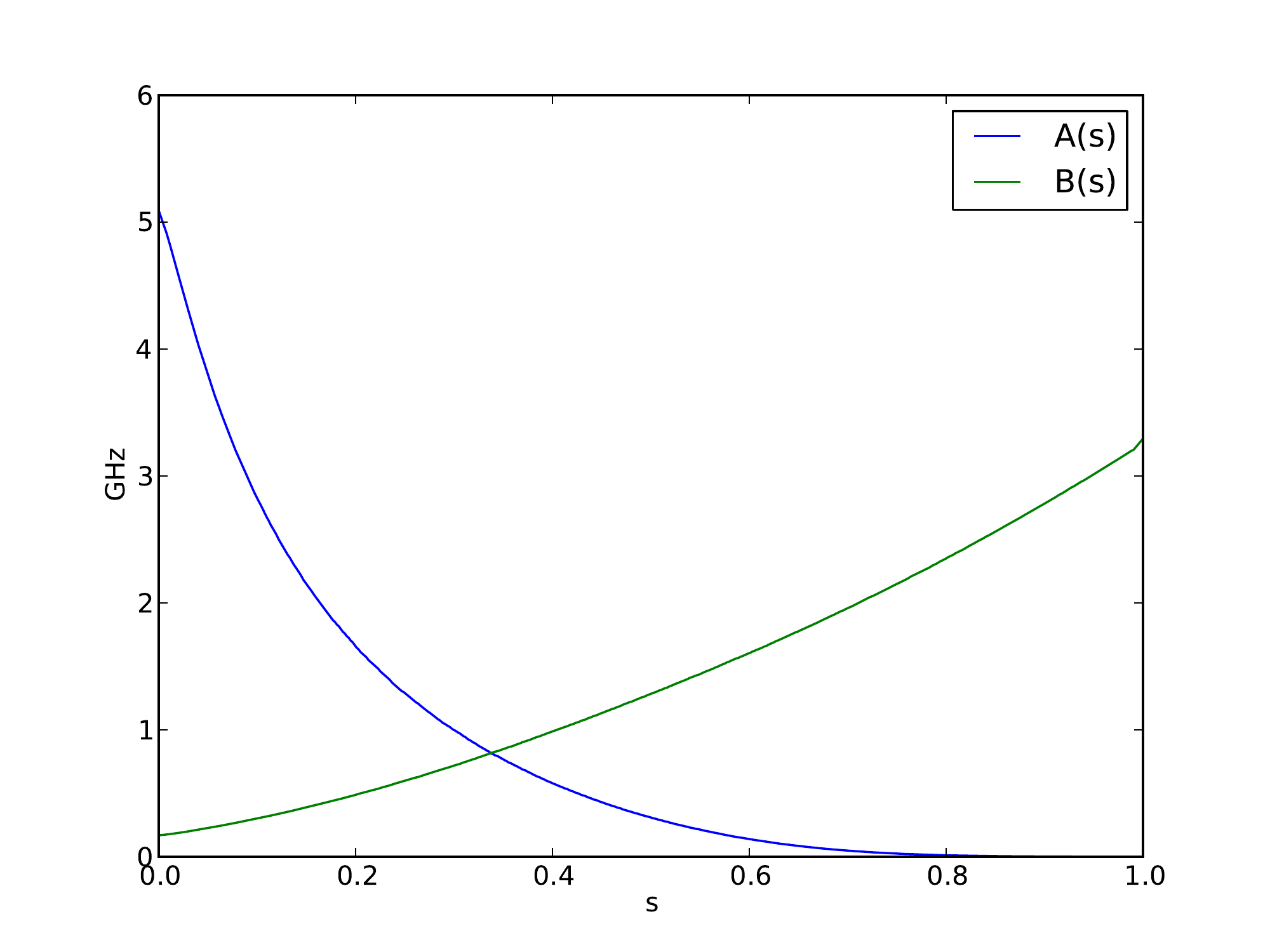}
\caption{(Color online)
Quantum annealing schedules employed by the D-Wave processor, where
$s=t/t_f$.}
\label{fig:QASchedules}
\end{figure}

The D-Wave device implements the quantum annealing algorithm via
superconducting compound Josephson junction flux qubits
\cite{johnson:11}. The objective is to find the ground state of the
Ising problem Hamiltonian ${\mathcal H}_{\rm P}$ presented in
Eq.~\eqref{eq:ham} defined on the D-Wave \emph{Chimera} graph; see
Fig.~\ref{fig:chimera}. This is attempted by applying and slowly
removing a transverse field. The time-dependent Hamiltonian is thus
given by
\begin{equation}
{\mathcal H}(s) = A(s) {\mathcal H}_{\rm D} + B(s) {\mathcal H}_{\rm P} , 
\end{equation}
where the \emph{driver} Hamiltonian $ {\mathcal H}_D = \sum_i
\sigma^{x}_i$, $s \in [0,1] $, and $A(s), B(s)$, which control the
relative magnitudes of driver and problem Hamiltonians are,
respectively, decreasing and increasing in $s$. Plots of $A(s)$ and
$B(s)$ are shown in Fig.~\ref{fig:QASchedules}. The parameter $s$ can be
translated to time $t$ via the relation $t= s t_f$, where $t_f$ is the
annealing time.

\subsection{D-Wave Two Methodology}

An annealing time of $20 \mu s$ is used for all experimental runs on
the DW2 processor, which is cooled to a temperature of $18$mK. Each problem
instance is run $N_{\rm R} = 10^4$ times in $N_{\rm G} = 10$ batches of
randomly-chosen gauge transformations in order to provide 
protection against parameter noise and control errors.  To generate a
gauge transformation, a set of $N$ random variables $\{t_i\}$, with 
$t_i \in \{-1,1\}$, is sampled uniformly, and the transformation
\begin{equation}
h'_i \gets h_i t_i 
\;\;\;\;\;\;\;\;\;\;
J'_{ij} \gets J_{ij} t_i t_j
\end{equation}
is made. In principle, this procedure does not fundamentally change the
problem, but due to parameter noise on the physical device, each gauge
transformation of a given instance will, in reality, correspond to a
different Hamiltonian.

Following the analysis performed in Ref.~\cite{boixo:14}, an instance's
success probability across gauges is derived from the geometric mean of
the gauges' failure rates. If $p_g$ is the observed success probability
of a gauge $g$, then
\begin{equation}
\bar{p} = 1-\prod_{g=1}^{N_{\rm G}}( 1-p_g)^{1/N_{\rm G}} .
\end{equation}
A ``success'' is defined as the occurrence of a state meeting a
criterion, for example, of having ground-state energy $E_0$, or with
energy lying in a range $[E_0, E_0+\Delta]$, $\Delta > 0$, of the
minimum.

The DW2 device is run in the so-called ``autoscaling'' mode for all
problems, which adjusts the nominally specified $J$ and $h$ parameters
to fully use the range allowed by the device.

\subsection{Simulated annealing methodology}

For the software-based simulated annealing experiments, we use the
codes developed by Isakov {\em et al.}~\cite{isakov:15} to ensure a
fair comparison with previous studies. The authors present a variant of
SA that exploits the bipartite nature of topologies such as the Chimera
graph's in order to halve the number of variables being simulated. This
optimization results in considerably improved performance over plain SA.
In this study we use the \texttt{an\_ss\_ge\_nf\_bp\_vdeg} routine.

All instances are simulated $N_{\rm R} = 10^4$ times for $N_{\rm
sw}=900$ Monte Carlo sweeps each; clearly, no advantage would be gained
from gauge transformations in the software case. The default geometric
annealing schedule described in Ref.~\cite{isakov:15} was adequate for
our purposes, but the (inverse) temperature scales were appropriately
adjusted for each instance class. The parameters of the simulation are
listed in Table \ref{tab:SAParams}.

Note that we choose  $N_{\rm sw}=900$, such that the average success
probabilities for the DW2 device agree with the SA simulations for the
commonly studied bimodal (U$_1$) disorder. We choose this approach to
provide a baseline for all other instance classes.  Simulations with
$N_{\rm sw}=2000$ sweeps showed qualitatively similar results.

\begin{table}
\caption{
Simulated annealing parameters used for the different instance classes.
For each type of disorder class $J_{ij}$, $N_{\rm sw}$ Monte Carlo
sweeps are performed on an annealing schedule from $\beta_{\rm i}$
to $\beta_{\rm f}$.
\label{tab:SAParams}}
\begin{tabular*}{\columnwidth}{@{\extracolsep{\fill}} l l c c c}
\hline
\hline
Class & $J_{ij}$ & $N_{\rm sw}$ & $\beta_{\rm i}$ & $\beta_{\rm f}$ \\
\hline
U$_1$    & $\{\pm 1\}$                      & $900$ & $0.1000$ & $3.000$ \\
U$_4$    & $\{\pm 1,\pm 2,\pm 3,\pm 4\}$    & $900$ & $0.2500$ & $7.500$ \\
S$_{28}$ & $\{\pm 8,\pm 13,\pm 19,\pm 28\}$ & $900$ & $0.0357$ & $1.071$ \\
\hline
\hline
\end{tabular*}
\end{table}

\section{Parallel tempering Monte Carlo simulation details}
\label{app:ptmc}

To compute the overlap distribution $P(q)$ we perform finite-temperature
parallel tempering Monte Carlo simulations
\cite{hukushima:96,geyer:91,katzgraber:06a} combined with isoenergetic
cluster moves \cite{zhu:15b} to speed up the simulations. We choose a
temperature set with $30$ temperatures and the lowest temperature
$T_{\rm min} = 0.212$ is chosen such that thermalization can be
completed in a meaningful time and features in the overlap distribution
are well defined. Two replicas with $N = 496$ spins and the same
disorder are thermalized for $2^{23}$ Monte Carlo sweeps and $P(q)$ is
measured over an additional $2^{23}$ Monte Carlo sweeps to obtain
high-resolution data. We compute $10^5$ randomly chosen disorder
instances for each problem class. The data are then mined according to
predefined criteria (see Sec.~\ref{sec:tune}).

\section{Experimental Results}
\label{app:results}

Table \ref{tab:res} lists the numerical values of the average success
probabilities for the different instance classes we study either on the
DW2 device or with SA codes. All numbers are averaged via a jackknife
procedure over $N_{\rm sa}$ instances of the disorder.

\begin{table}
\caption{
Raw results of the experiments on the DW2 device and the
simulated annealing (SA) codes for the different instance classes
we study.  Listed are average success probabilities ($p_{\rm av}$) in
percent, as well as the number of disorder instances $N_{\rm sa}$
studied.
\label{tab:res}}
\begin{tabular*}{\columnwidth}{@{\extracolsep{\fill}} l l l l}
\hline
\hline
Class & $N_{\rm sa}$ & $p_{\rm av}$(\%) [DW2] & $p_{\rm av}$(\%) [SA] \\
\hline
S$_{28}$ (thick barriers) & $2239$ & $0.032(2)$ & $0.141(3)$  \\
S$_{28}$ (thin barriers)  & $1816$ & $0.054(4)$ & $0.140(3)$  \\
S$_{28}$ (small barriers) & $2637$ & $0.095(4)$ & $0.246(5)$  \\
\hline
U$_4$                     & $2000$ & $0.50(3)$  & $0.243(5)$  \\
\hline
U$_1$                     & $2000$ & $7.24(26)$ & $7.49(14)$ \\
\hline
\hline
\end{tabular*}
\end{table}

\section{Other Instance Classes Studied}
\label{app:other}

We also perform other experiments with different instance
classes.  However, these are either too easy or it is extremely
difficult to obtain unique ground-state instances. Note that for the
J$_4$ instances \cite{katzgraber:14}, where the interactions are
bimodally distributed and the bonds in the K$_{4,4}$ cells are a $1/4$,
as well as the S$_{1,3,7}$ small Sidon instances, we limit the number
of configurations that minimize the Hamiltonian to less than $32$
because too few unique ground states could be found. As such, we are
merely mentioning here the results to prevent other researchers from
attempting to study these systems. Average success probabilities are
listed in Table \ref{tab:extra}.

\begin{table}
\caption{
Raw results of the experiments on the DW2 device and the
SA codes for additional instance classes
we study.  Listed are average success probabilities ($p_{\rm av}$) in
percent, as well as the number of disorder instances $N_{\rm sa}$
studied.
\label{tab:extra}}
\begin{tabular*}{\columnwidth}{@{\extracolsep{\fill}} l l l l}
\hline
\hline
Class & $N_{\rm sa}$ & $p_{\rm av}$(\%) [DW2] & $p_{\rm av}$(\%) [SA] \\
\hline
\hline
J$_4$ (thick barriers)    	& $1250$ & $0.50(3)$   & $4.1(1)$   \\
J$_4$ (small barriers)    	& $2035$ & $1.96(6)$   & $13.3(2)$   \\
\hline
S$_{1,3,7}$ (thick barriers) 	& $1615$ & $0.063(4)$  & $0.59(1)$ \\
S$_{1,3,7}$ (small barriers)    & $1582$ & $0.22(1)$   & $1.14(2)$   \\
\hline
\hline
\end{tabular*}
\end{table}

\bibliography{refs,comments}

\begin{thebibliography}{108}
\expandafter\ifx\csname natexlab\endcsname\relax\def\natexlab#1{#1}\fi
\expandafter\ifx\csname bibnamefont\endcsname\relax
  \def\bibnamefont#1{#1}\fi
\expandafter\ifx\csname bibfnamefont\endcsname\relax
  \def\bibfnamefont#1{#1}\fi
\expandafter\ifx\csname citenamefont\endcsname\relax
  \def\citenamefont#1{#1}\fi
\expandafter\ifx\csname url\endcsname\relax
  \def\url#1{\texttt{#1}}\fi
\expandafter\ifx\csname urlprefix\endcsname\relax\def\urlprefix{URL }\fi
\providecommand{\bibinfo}[2]{#2}
\providecommand{\eprint}[2][]{\url{#2}}

\bibitem[{\citenamefont{Hartmann and Rieger}(2001)}]{hartmann:01}
\bibinfo{author}{\bibfnamefont{A.~K.} \bibnamefont{Hartmann}} \bibnamefont{and}
  \bibinfo{author}{\bibfnamefont{H.}~\bibnamefont{Rieger}},
  \emph{\bibinfo{title}{Optimization Algorithms in Physics}}
  (\bibinfo{publisher}{Wiley-VCH}, \bibinfo{address}{Berlin},
  \bibinfo{year}{2001}).

\bibitem[{jue(2001)}]{juenger:01}
\emph{\bibinfo{title}{Lecture notes in computer science 2241}}, in
  \emph{\bibinfo{booktitle}{Computational Combinatorial Optimization}}, edited
  by \bibinfo{editor}{\bibfnamefont{M.}~\bibnamefont{J{\"u}nger}}
  \bibnamefont{and} \bibinfo{editor}{\bibfnamefont{D.}~\bibnamefont{Naddef}}
  (\bibinfo{publisher}{Springer Verlag}, \bibinfo{address}{Heidelberg},
  \bibinfo{year}{2001}), vol. \bibinfo{volume}{2241}.

\bibitem[{\citenamefont{Hartmann and Rieger}(2004)}]{hartmann:04}
\bibinfo{author}{\bibfnamefont{A.~K.} \bibnamefont{Hartmann}} \bibnamefont{and}
  \bibinfo{author}{\bibfnamefont{H.}~\bibnamefont{Rieger}},
  \emph{\bibinfo{title}{New Optimization Algorithms in Physics}}
  (\bibinfo{publisher}{Wiley-VCH}, \bibinfo{address}{Berlin},
  \bibinfo{year}{2004}).

\bibitem[{\citenamefont{Moore}(1965)}]{moore:65}
\bibinfo{author}{\bibfnamefont{G.}~\bibnamefont{Moore}},
  \emph{\bibinfo{title}{{{Cramming more components onto integrated
  circuits}}}}, \bibinfo{journal}{Electronics Magazine}
  \textbf{\bibinfo{volume}{38}}, \bibinfo{pages}{114} (\bibinfo{year}{1965}).

\bibitem[{\citenamefont{Nielsen and Chuang}(2000)}]{nielsen:00}
\bibinfo{author}{\bibfnamefont{M.~A.} \bibnamefont{Nielsen}} \bibnamefont{and}
  \bibinfo{author}{\bibfnamefont{I.~L.} \bibnamefont{Chuang}},
  \emph{\bibinfo{title}{Quantum Computation and Quantum Information}}
  (\bibinfo{publisher}{{Cambridge University Press}},
  \bibinfo{address}{Cambridge}, \bibinfo{year}{2000}).

\bibitem[{\citenamefont{Nishimori}(2001)}]{nishimori:01}
\bibinfo{author}{\bibfnamefont{H.}~\bibnamefont{Nishimori}},
  \emph{\bibinfo{title}{{Statistical Physics of Spin Glasses and Information
  Processing: An Introduction}}} (\bibinfo{publisher}{Oxford University Press},
  \bibinfo{address}{New York}, \bibinfo{year}{2001}).

\bibitem[{\citenamefont{Finnila et~al.}(1994)\citenamefont{Finnila, Gomez,
  Sebenik, Stenson, and Doll}}]{finnila:94}
\bibinfo{author}{\bibfnamefont{A.~B.} \bibnamefont{Finnila}},
  \bibinfo{author}{\bibfnamefont{M.~A.} \bibnamefont{Gomez}},
  \bibinfo{author}{\bibfnamefont{C.}~\bibnamefont{Sebenik}},
  \bibinfo{author}{\bibfnamefont{C.}~\bibnamefont{Stenson}}, \bibnamefont{and}
  \bibinfo{author}{\bibfnamefont{J.~D.} \bibnamefont{Doll}},
  \emph{\bibinfo{title}{{{Quantum annealing: A new method for minimizing
  multidimensional functions}}}}, \bibinfo{journal}{Chem. Phys. Lett.}
  \textbf{\bibinfo{volume}{219}}, \bibinfo{pages}{343} (\bibinfo{year}{1994}).

\bibitem[{\citenamefont{Kadowaki and Nishimori}(1998)}]{kadowaki:98}
\bibinfo{author}{\bibfnamefont{T.}~\bibnamefont{Kadowaki}} \bibnamefont{and}
  \bibinfo{author}{\bibfnamefont{H.}~\bibnamefont{Nishimori}},
  \emph{\bibinfo{title}{{{Quantum annealing in the transverse Ising model}}}},
  \bibinfo{journal}{Phys. Rev. E} \textbf{\bibinfo{volume}{58}},
  \bibinfo{pages}{5355} (\bibinfo{year}{1998}).

\bibitem[{\citenamefont{Brooke et~al.}(1999)\citenamefont{Brooke, Bitko,
  Rosenbaum, and Aepli}}]{brooke:99}
\bibinfo{author}{\bibfnamefont{J.}~\bibnamefont{Brooke}},
  \bibinfo{author}{\bibfnamefont{D.}~\bibnamefont{Bitko}},
  \bibinfo{author}{\bibfnamefont{T.~F.} \bibnamefont{Rosenbaum}},
  \bibnamefont{and} \bibinfo{author}{\bibfnamefont{G.}~\bibnamefont{Aepli}},
  \emph{\bibinfo{title}{Quantum annealing of a disordered magnet}},
  \bibinfo{journal}{Science} \textbf{\bibinfo{volume}{284}},
  \bibinfo{pages}{779} (\bibinfo{year}{1999}).

\bibitem[{\citenamefont{{Farhi} et~al.}(2000)\citenamefont{{Farhi},
  {Goldstone}, {Gutmann}, and {Sipser}}}]{farhi:00}
\bibinfo{author}{\bibfnamefont{E.}~\bibnamefont{{Farhi}}},
  \bibinfo{author}{\bibfnamefont{J.}~\bibnamefont{{Goldstone}}},
  \bibinfo{author}{\bibfnamefont{S.}~\bibnamefont{{Gutmann}}},
  \bibnamefont{and} \bibinfo{author}{\bibfnamefont{M.}~\bibnamefont{{Sipser}}},
  \emph{\bibinfo{title}{{{Quantum Computation by Adiabatic Evolution}}}}
  (\bibinfo{year}{2000}), \bibinfo{note}{arXiv:quant-ph/0001106}.

\bibitem[{\citenamefont{Roland and Cerf}(2002)}]{roland:02}
\bibinfo{author}{\bibfnamefont{J.}~\bibnamefont{Roland}} \bibnamefont{and}
  \bibinfo{author}{\bibfnamefont{N.~J.} \bibnamefont{Cerf}},
  \emph{\bibinfo{title}{{{Quantum search by local adiabatic evolution}}}},
  \bibinfo{journal}{Phys. Rev. A} \textbf{\bibinfo{volume}{65}},
  \bibinfo{pages}{042308} (\bibinfo{year}{2002}).

\bibitem[{\citenamefont{Santoro et~al.}(2002)\citenamefont{Santoro,
  Marto\v{n}\'ak, and Car}}]{santoro:02}
\bibinfo{author}{\bibfnamefont{G.}~\bibnamefont{Santoro}},
  \bibinfo{author}{\bibfnamefont{E.}~\bibnamefont{Marto\v{n}\'ak},
  \bibfnamefont{R.~Tosatti}}, \bibnamefont{and}
  \bibinfo{author}{\bibfnamefont{R.}~\bibnamefont{Car}},
  \emph{\bibinfo{title}{Theory of quantum annealing of an {I}sing spin glass}},
  \bibinfo{journal}{Science} \textbf{\bibinfo{volume}{295}},
  \bibinfo{pages}{2427} (\bibinfo{year}{2002}).

\bibitem[{\citenamefont{Das and Chakrabarti}(2005)}]{das:05}
\bibinfo{author}{\bibfnamefont{A.}~\bibnamefont{Das}} \bibnamefont{and}
  \bibinfo{author}{\bibfnamefont{B.~K.} \bibnamefont{Chakrabarti}},
  \emph{\bibinfo{title}{{{Quantum Annealing and Related Optimization
  Methods}}}} (\bibinfo{publisher}{Edited by A.~Das and B.K.~Chakrabarti,
  Lecture Notes in Physics 679, Berlin: Springer}, \bibinfo{year}{2005}).

\bibitem[{\citenamefont{Santoro and Tosatti}(2006)}]{santoro:06}
\bibinfo{author}{\bibfnamefont{G.~E.} \bibnamefont{Santoro}} \bibnamefont{and}
  \bibinfo{author}{\bibfnamefont{E.}~\bibnamefont{Tosatti}},
  \emph{\bibinfo{title}{{{TOPICAL REVIEW: Optimization using quantum mechanics:
  quantum annealing through adiabatic evolution}}}}, \bibinfo{journal}{J. Phys.
  A} \textbf{\bibinfo{volume}{39}}, \bibinfo{pages}{R393}
  (\bibinfo{year}{2006}).

\bibitem[{\citenamefont{Lidar}(2008)}]{lidar:08}
\bibinfo{author}{\bibfnamefont{D.~A.} \bibnamefont{Lidar}},
  \emph{\bibinfo{title}{{{Towards Fault Tolerant Adiabatic Quantum
  Computation}}}}, \bibinfo{journal}{Phys. Rev. Lett.}
  \textbf{\bibinfo{volume}{100}}, \bibinfo{pages}{160506}
  (\bibinfo{year}{2008}).

\bibitem[{\citenamefont{Das and Chakrabarti}(2008)}]{das:08}
\bibinfo{author}{\bibfnamefont{A.}~\bibnamefont{Das}} \bibnamefont{and}
  \bibinfo{author}{\bibfnamefont{B.~K.} \bibnamefont{Chakrabarti}},
  \emph{\bibinfo{title}{{{Quantum Annealing and Analog Quantum Computation}}}},
  \bibinfo{journal}{Rev. Mod. Phys.} \textbf{\bibinfo{volume}{80}},
  \bibinfo{pages}{1061} (\bibinfo{year}{2008}).

\bibitem[{\citenamefont{Morita and Nishimori}(2008)}]{morita:08}
\bibinfo{author}{\bibfnamefont{S.}~\bibnamefont{Morita}} \bibnamefont{and}
  \bibinfo{author}{\bibfnamefont{H.}~\bibnamefont{Nishimori}},
  \emph{\bibinfo{title}{{{Mathematical Foundation of Quantum Annealing}}}},
  \bibinfo{journal}{J. Math. Phys.} \textbf{\bibinfo{volume}{49}},
  \bibinfo{pages}{125210} (\bibinfo{year}{2008}).

\bibitem[{\citenamefont{Mukherjee and Chakrabarti}(2015)}]{mukherjee:15}
\bibinfo{author}{\bibfnamefont{S.}~\bibnamefont{Mukherjee}} \bibnamefont{and}
  \bibinfo{author}{\bibfnamefont{B.~K.} \bibnamefont{Chakrabarti}},
  \emph{\bibinfo{title}{{{Multivariable optimization: Quantum annealing and
  computation}}}}, \bibinfo{journal}{Eur. Phys. J. Special Topics}
  \textbf{\bibinfo{volume}{224}}, \bibinfo{pages}{17} (\bibinfo{year}{2015}).

\bibitem[{\citenamefont{Johnson et~al.}(2011)\citenamefont{Johnson, Amin,
  Gildert, Lanting, Hamze, Dickson, Harris, Berkley, Johansson, Bunyk
  et~al.}}]{johnson:11}
\bibinfo{author}{\bibfnamefont{M.~W.} \bibnamefont{Johnson}},
  \bibinfo{author}{\bibfnamefont{M.~H.~S.} \bibnamefont{Amin}},
  \bibinfo{author}{\bibfnamefont{S.}~\bibnamefont{Gildert}},
  \bibinfo{author}{\bibfnamefont{T.}~\bibnamefont{Lanting}},
  \bibinfo{author}{\bibfnamefont{F.}~\bibnamefont{Hamze}},
  \bibinfo{author}{\bibfnamefont{N.}~\bibnamefont{Dickson}},
  \bibinfo{author}{\bibfnamefont{R.}~\bibnamefont{Harris}},
  \bibinfo{author}{\bibfnamefont{A.~J.} \bibnamefont{Berkley}},
  \bibinfo{author}{\bibfnamefont{J.}~\bibnamefont{Johansson}},
  \bibinfo{author}{\bibfnamefont{P.}~\bibnamefont{Bunyk}},
  \bibnamefont{et~al.}, \emph{\bibinfo{title}{{Quantum annealing with
  manufactured spins}}}, \bibinfo{journal}{Nature}
  \textbf{\bibinfo{volume}{473}}, \bibinfo{pages}{194} (\bibinfo{year}{2011}).

\bibitem[{\citenamefont{Amin and Choi}(2009)}]{amin:09}
\bibinfo{author}{\bibfnamefont{M.~H.~S.} \bibnamefont{Amin}} \bibnamefont{and}
  \bibinfo{author}{\bibfnamefont{V.}~\bibnamefont{Choi}},
  \emph{\bibinfo{title}{{{First-order quantum phase transition in adiabatic
  quantum computation}}}}, \bibinfo{journal}{Phys. Rev. A}
  \textbf{\bibinfo{volume}{80}}, \bibinfo{pages}{062326}
  (\bibinfo{year}{2009}).

\bibitem[{\citenamefont{Young et~al.}(2010)\citenamefont{Young, Knysh, and
  Smelyanskiy}}]{young:10}
\bibinfo{author}{\bibfnamefont{A.~P.} \bibnamefont{Young}},
  \bibinfo{author}{\bibfnamefont{S.}~\bibnamefont{Knysh}}, \bibnamefont{and}
  \bibinfo{author}{\bibfnamefont{V.~N.} \bibnamefont{Smelyanskiy}},
  \emph{\bibinfo{title}{{{First-Order Phase Transition in the Quantum Adiabatic
  Algorithm}}}}, \bibinfo{journal}{Phys. Rev. Lett.}
  \textbf{\bibinfo{volume}{104}}, \bibinfo{pages}{020502}
  (\bibinfo{year}{2010}).

\bibitem[{\citenamefont{Hen and Young}(2011)}]{hen:11}
\bibinfo{author}{\bibfnamefont{I.}~\bibnamefont{Hen}} \bibnamefont{and}
  \bibinfo{author}{\bibfnamefont{A.~P.} \bibnamefont{Young}},
  \emph{\bibinfo{title}{{{Exponential complexity of the quantum adiabatic
  algorithm for certain satisfiability problems}}}}, \bibinfo{journal}{Phys.
  Rev. E} \textbf{\bibinfo{volume}{84}}, \bibinfo{pages}{061152}
  (\bibinfo{year}{2011}).

\bibitem[{\citenamefont{Matsuda et~al.}(2009)\citenamefont{Matsuda, Nishimori,
  and Katzgraber}}]{matsuda:09}
\bibinfo{author}{\bibfnamefont{Y.}~\bibnamefont{Matsuda}},
  \bibinfo{author}{\bibfnamefont{H.}~\bibnamefont{Nishimori}},
  \bibnamefont{and} \bibinfo{author}{\bibfnamefont{H.~G.}
  \bibnamefont{Katzgraber}}, \emph{\bibinfo{title}{{Ground-state statistics
  from annealing algorithms: quantum versus classical approaches}}},
  \bibinfo{journal}{New J. Phys.} \textbf{\bibinfo{volume}{11}},
  \bibinfo{pages}{073021} (\bibinfo{year}{2009}).

\bibitem[{\citenamefont{Kirkpatrick et~al.}(1983)\citenamefont{Kirkpatrick,
  {Gelatt, Jr.}, and Vecchi}}]{kirkpatrick:83}
\bibinfo{author}{\bibfnamefont{S.}~\bibnamefont{Kirkpatrick}},
  \bibinfo{author}{\bibfnamefont{C.~D.} \bibnamefont{{Gelatt, Jr.}}},
  \bibnamefont{and} \bibinfo{author}{\bibfnamefont{M.~P.}
  \bibnamefont{Vecchi}}, \emph{\bibinfo{title}{Optimization by simulated
  annealing}}, \bibinfo{journal}{Science} \textbf{\bibinfo{volume}{220}},
  \bibinfo{pages}{671} (\bibinfo{year}{1983}).

\bibitem[{\citenamefont{Binder and Young}(1986)}]{binder:86}
\bibinfo{author}{\bibfnamefont{K.}~\bibnamefont{Binder}} \bibnamefont{and}
  \bibinfo{author}{\bibfnamefont{A.~P.} \bibnamefont{Young}},
  \emph{\bibinfo{title}{Spin glasses: Experimental facts, theoretical concepts
  and open questions}}, \bibinfo{journal}{Rev. Mod. Phys.}
  \textbf{\bibinfo{volume}{58}}, \bibinfo{pages}{801} (\bibinfo{year}{1986}).

\bibitem[{\citenamefont{Stein and Newman}(2013)}]{stein:13}
\bibinfo{author}{\bibfnamefont{D.~L.} \bibnamefont{Stein}} \bibnamefont{and}
  \bibinfo{author}{\bibfnamefont{C.~M.} \bibnamefont{Newman}},
  \emph{\bibinfo{title}{{Spin Glasses and Complexity}}}, Primers in Complex
  Systems (\bibinfo{publisher}{Princeton University Press},
  \bibinfo{year}{2013}).

\bibitem[{\citenamefont{Lucas}(2014)}]{lucas:14}
\bibinfo{author}{\bibfnamefont{A.}~\bibnamefont{Lucas}},
  \emph{\bibinfo{title}{{Ising formulations of many NP problems}}},
  \bibinfo{journal}{Front. Physics} \textbf{\bibinfo{volume}{12}},
  \bibinfo{pages}{5} (\bibinfo{year}{2014}).

\bibitem[{\citenamefont{{R{\o}nnow} et~al.}(2014)\citenamefont{{R{\o}nnow},
  {Wang}, {Job}, {Boixo}, {Isakov}, {Wecker}, {Martinis}, {Lidar}, and
  {Troyer}}}]{ronnow:14a}
\bibinfo{author}{\bibfnamefont{T.~F.} \bibnamefont{{R{\o}nnow}}},
  \bibinfo{author}{\bibfnamefont{Z.}~\bibnamefont{{Wang}}},
  \bibinfo{author}{\bibfnamefont{J.}~\bibnamefont{{Job}}},
  \bibinfo{author}{\bibfnamefont{S.}~\bibnamefont{{Boixo}}},
  \bibinfo{author}{\bibfnamefont{S.~V.} \bibnamefont{{Isakov}}},
  \bibinfo{author}{\bibfnamefont{D.}~\bibnamefont{{Wecker}}},
  \bibinfo{author}{\bibfnamefont{J.~M.} \bibnamefont{{Martinis}}},
  \bibinfo{author}{\bibfnamefont{D.~A.} \bibnamefont{{Lidar}}},
  \bibnamefont{and} \bibinfo{author}{\bibfnamefont{M.}~\bibnamefont{{Troyer}}},
  \emph{\bibinfo{title}{{Defining and detecting quantum speedup}}},
  \bibinfo{journal}{Science} \textbf{\bibinfo{volume}{345}}
  (\bibinfo{year}{2014}).

\bibitem[{com({\natexlab{a}})}]{comment:limited}
\bibinfo{note}{For the sake of brevity we shall refer to ``{\em limited quantum
  speedup}'' simply as ``{\em quantum speedup}.''}.

\bibitem[{\citenamefont{Somma et~al.}(2012)\citenamefont{Somma, Nagaj, and
  Kieferov\'a}}]{nagaj:12}
\bibinfo{author}{\bibfnamefont{R.~D.} \bibnamefont{Somma}},
  \bibinfo{author}{\bibfnamefont{D.}~\bibnamefont{Nagaj}}, \bibnamefont{and}
  \bibinfo{author}{\bibfnamefont{M.}~\bibnamefont{Kieferov\'a}},
  \emph{\bibinfo{title}{{{Quantum Speedup by Quantum Annealing}}}},
  \bibinfo{journal}{Phys. Rev. Lett.} \textbf{\bibinfo{volume}{109}},
  \bibinfo{pages}{050501} (\bibinfo{year}{2012}).

\bibitem[{\citenamefont{Pudenz et~al.}(2014)\citenamefont{Pudenz, Albash, and
  Lidar}}]{pudenz:13}
\bibinfo{author}{\bibfnamefont{K.~L.} \bibnamefont{Pudenz}},
  \bibinfo{author}{\bibfnamefont{T.}~\bibnamefont{Albash}}, \bibnamefont{and}
  \bibinfo{author}{\bibfnamefont{D.~A.} \bibnamefont{Lidar}},
  \emph{\bibinfo{title}{Error-corrected quantum annealing with hundreds of
  qubits}}, \bibinfo{journal}{Nat. Commun.} \textbf{\bibinfo{volume}{5}},
  \bibinfo{pages}{3243} (\bibinfo{year}{2014}).

\bibitem[{\citenamefont{{Boixo} et~al.}(2013)\citenamefont{{Boixo}, {Albash},
  {Spedalieri}, {Chancellor}, and {Lidar}}}]{boixo:13a}
\bibinfo{author}{\bibfnamefont{S.}~\bibnamefont{{Boixo}}},
  \bibinfo{author}{\bibfnamefont{T.}~\bibnamefont{{Albash}}},
  \bibinfo{author}{\bibfnamefont{F.~M.} \bibnamefont{{Spedalieri}}},
  \bibinfo{author}{\bibfnamefont{N.}~\bibnamefont{{Chancellor}}},
  \bibnamefont{and} \bibinfo{author}{\bibfnamefont{D.~A.}
  \bibnamefont{{Lidar}}}, \emph{\bibinfo{title}{{{Experimental signature of
  programmable quantum annealing}}}}, \bibinfo{journal}{Nat. Comm.}
  \textbf{\bibinfo{volume}{4}}, \bibinfo{pages}{2067} (\bibinfo{year}{2013}).

\bibitem[{\citenamefont{{Boixo}
  et~al.}(2014{\natexlab{a}})\citenamefont{{Boixo}, {R{\o}nnow}, {Isakov},
  {Wang}, {Wecker}, {Lidar}, {Martinis}, and {Troyer}}}]{boixo:14}
\bibinfo{author}{\bibfnamefont{S.}~\bibnamefont{{Boixo}}},
  \bibinfo{author}{\bibfnamefont{T.~F.} \bibnamefont{{R{\o}nnow}}},
  \bibinfo{author}{\bibfnamefont{S.~V.} \bibnamefont{{Isakov}}},
  \bibinfo{author}{\bibfnamefont{Z.}~\bibnamefont{{Wang}}},
  \bibinfo{author}{\bibfnamefont{D.}~\bibnamefont{{Wecker}}},
  \bibinfo{author}{\bibfnamefont{D.~A.} \bibnamefont{{Lidar}}},
  \bibinfo{author}{\bibfnamefont{J.~M.} \bibnamefont{{Martinis}}},
  \bibnamefont{and} \bibinfo{author}{\bibfnamefont{M.}~\bibnamefont{{Troyer}}},
  \emph{\bibinfo{title}{{Evidence for quantum annealing with more than one
  hundred qubits}}}, \bibinfo{journal}{Nat. Phys,}
  \textbf{\bibinfo{volume}{10}}, \bibinfo{pages}{218}
  (\bibinfo{year}{2014}{\natexlab{a}}).

\bibitem[{\citenamefont{Katzgraber et~al.}(2014)\citenamefont{Katzgraber,
  Hamze, and Andrist}}]{katzgraber:14}
\bibinfo{author}{\bibfnamefont{H.~G.} \bibnamefont{Katzgraber}},
  \bibinfo{author}{\bibfnamefont{F.}~\bibnamefont{Hamze}}, \bibnamefont{and}
  \bibinfo{author}{\bibfnamefont{R.~S.} \bibnamefont{Andrist}},
  \emph{\bibinfo{title}{{Glassy Chimeras Could Be Blind to Quantum Speedup:
  Designing Better Benchmarks for Quantum Annealing Machines}}},
  \bibinfo{journal}{Phys. Rev. X} \textbf{\bibinfo{volume}{4}},
  \bibinfo{pages}{021008} (\bibinfo{year}{2014}).

\bibitem[{\citenamefont{Bunyk}(2014)}]{bunyk:14}
\bibinfo{author}{\bibfnamefont{P.}~\bibnamefont{Bunyk}},
  \emph{\bibinfo{title}{{Architectural Considerations in the Design of a
  Superconducting Quantum Annealing Processor}}}, \bibinfo{journal}{IEEE Trans.
  Appl. Supercond.} \textbf{\bibinfo{volume}{24}}, \bibinfo{pages}{1}
  (\bibinfo{year}{2014}).

\bibitem[{\citenamefont{Hen et~al.}(2015)\citenamefont{Hen, Job, Albash,
  {R{\o}nnow}, Troyer, and Lidar}}]{hen:15}
\bibinfo{author}{\bibfnamefont{I.}~\bibnamefont{Hen}},
  \bibinfo{author}{\bibfnamefont{J.}~\bibnamefont{Job}},
  \bibinfo{author}{\bibfnamefont{T.}~\bibnamefont{Albash}},
  \bibinfo{author}{\bibfnamefont{T.~F.} \bibnamefont{{R{\o}nnow}}},
  \bibinfo{author}{\bibfnamefont{M.}~\bibnamefont{Troyer}}, \bibnamefont{and}
  \bibinfo{author}{\bibfnamefont{D.}~\bibnamefont{Lidar}},
  \emph{\bibinfo{title}{{Probing for quantum speedup in spin glass problems
  with planted solutions}}} (\bibinfo{year}{2015}),
  \bibinfo{note}{(arXiv:quant-phys/1502.01663)}.

\bibitem[{\citenamefont{{Venturelli} et~al.}(2014)\citenamefont{{Venturelli},
  {Mandr{\`a}}, {Knysh}, {O'Gorman}, {Biswas}, and
  {Smelyanskiy}}}]{venturelli:15}
\bibinfo{author}{\bibfnamefont{D.}~\bibnamefont{{Venturelli}}},
  \bibinfo{author}{\bibfnamefont{S.}~\bibnamefont{{Mandr{\`a}}}},
  \bibinfo{author}{\bibfnamefont{S.}~\bibnamefont{{Knysh}}},
  \bibinfo{author}{\bibfnamefont{B.}~\bibnamefont{{O'Gorman}}},
  \bibinfo{author}{\bibfnamefont{R.}~\bibnamefont{{Biswas}}}, \bibnamefont{and}
  \bibinfo{author}{\bibfnamefont{V.}~\bibnamefont{{Smelyanskiy}}},
  \emph{\bibinfo{title}{{{Quantum Optimization of Fully-Connected Spin
  Glasses}}}} (\bibinfo{year}{2014}),
  \bibinfo{note}{(arXiv:cond-mat/1406.7553)}.

\bibitem[{\citenamefont{Zhu et~al.}(2015)\citenamefont{Zhu, Ochoa, Hamze,
  Schnabel, and Katzgraber}}]{zhu:15a}
\bibinfo{author}{\bibfnamefont{Z.}~\bibnamefont{Zhu}},
  \bibinfo{author}{\bibfnamefont{A.~J.} \bibnamefont{Ochoa}},
  \bibinfo{author}{\bibfnamefont{F.}~\bibnamefont{Hamze}},
  \bibinfo{author}{\bibfnamefont{S.}~\bibnamefont{Schnabel}}, \bibnamefont{and}
  \bibinfo{author}{\bibfnamefont{H.~G.} \bibnamefont{Katzgraber}},
  \emph{\bibinfo{title}{{{Best-case performance of quantum annealers on native
  spin-glass benchmarks: How chaos can affect success probabilities}}}}
  (\bibinfo{year}{2015}), \bibinfo{note}{(arXiv:1505.02278)}.

\bibitem[{\citenamefont{{Isakov} et~al.}(2015)\citenamefont{{Isakov},
  {Zintchenko}, {R{\o}nnow}, and {Troyer}}}]{isakov:15}
\bibinfo{author}{\bibfnamefont{S.~V.} \bibnamefont{{Isakov}}},
  \bibinfo{author}{\bibfnamefont{I.~N.} \bibnamefont{{Zintchenko}}},
  \bibinfo{author}{\bibfnamefont{T.~F.} \bibnamefont{{R{\o}nnow}}},
  \bibnamefont{and} \bibinfo{author}{\bibfnamefont{M.}~\bibnamefont{{Troyer}}},
  \emph{\bibinfo{title}{{{Optimized simulated annealing for Ising spin
  glasses}}}}, \bibinfo{journal}{Comput. Phys. Commun.}
  \textbf{\bibinfo{volume}{192}}, \bibinfo{pages}{265} (\bibinfo{year}{2015}),
  \bibinfo{note}{(see also ancillary material to arxiv:cond-mat/1401.1084)}.

\bibitem[{\citenamefont{Yucesoy et~al.}(2013)\citenamefont{Yucesoy, Machta, and
  Katzgraber}}]{yucesoy:13}
\bibinfo{author}{\bibfnamefont{B.}~\bibnamefont{Yucesoy}},
  \bibinfo{author}{\bibfnamefont{J.}~\bibnamefont{Machta}}, \bibnamefont{and}
  \bibinfo{author}{\bibfnamefont{H.~G.} \bibnamefont{Katzgraber}},
  \emph{\bibinfo{title}{{Correlations between the dynamics of parallel
  tempering and the free-energy landscape in spin glasses}}},
  \bibinfo{journal}{Phys. Rev. E} \textbf{\bibinfo{volume}{87}},
  \bibinfo{pages}{012104} (\bibinfo{year}{2013}).

\bibitem[{com({\natexlab{b}})}]{comment:pq}
\bibinfo{note}{We have performed \cite{melchert:15} detailed simulations to
  verify the conjecture that the spin-overlap distribution faithfully mirrors
  the free-energy landscape of a disordered system. Using both simple Monte
  Carlo \cite{katzgraber:09e} and improved extremal optimization
  \cite{middleton:04,boettcher:01} --- two methods that optimize in an
  inherently different way --- we study the algorithm's efficiency to traverse
  the energy landscape for the different Sidon instance classes studied. For
  both algorithms, when the spin-overlap distribution has a lot of structure
  (hard instances with thick and thin barriers), the Hamming distances between
  subsequent energy maxima and minima grow when approaching the ground-state
  energy. This means that the energy landscape is highly nontrivial in this
  case; i.e., deep valleys and tall mountains seem to exist. This is not the
  case for instances for which the spin overlap has a trivial structure
  reminiscent of a ferromagnet (easy instances). Here the Hamming distance as a
  function of the energy valley index decreases when the energies are close to
  the ground state, suggesting a featureless energy landscape.}

\bibitem[{com({\natexlab{c}})}]{comment:fss}
\bibinfo{note}{Figure 7 in Ref.~\cite{campbell:04} shows the change in energy
  when switching boundary conditions in two-dimensional Ising spin glasses on a
  square lattice. When the disorder is chosen from a discrete distribution (in
  this case bimodal), corrections to scaling are large and systems with more
  than approximately $4000$ spins are needed to truly probe the thermodynamic
  limit. This is also highlighted in the erratum to Ref.~\cite{katzgraber:14},
  Ref.~\cite{weigel:15}, where weak, but persistent corrections to scaling
  affected the determination of the critical exponents in the ferromagnetic
  sector.}

\bibitem[{\citenamefont{Campbell et~al.}(2004)\citenamefont{Campbell, Hartmann,
  and Katzgraber}}]{campbell:04}
\bibinfo{author}{\bibfnamefont{I.~A.} \bibnamefont{Campbell}},
  \bibinfo{author}{\bibfnamefont{A.~K.} \bibnamefont{Hartmann}},
  \bibnamefont{and} \bibinfo{author}{\bibfnamefont{H.~G.}
  \bibnamefont{Katzgraber}}, \emph{\bibinfo{title}{{Energy size effects of
  two-dimensional Ising spin glasses}}}, \bibinfo{journal}{Phys. Rev. B}
  \textbf{\bibinfo{volume}{70}}, \bibinfo{pages}{054429}
  (\bibinfo{year}{2004}).

\bibitem[{\citenamefont{Pudenz et~al.}(2015)\citenamefont{Pudenz, Albash, and
  Lidar}}]{pudenz:15}
\bibinfo{author}{\bibfnamefont{K.~L.} \bibnamefont{Pudenz}},
  \bibinfo{author}{\bibfnamefont{T.}~\bibnamefont{Albash}}, \bibnamefont{and}
  \bibinfo{author}{\bibfnamefont{D.~A.} \bibnamefont{Lidar}},
  \emph{\bibinfo{title}{{Quantum Annealing Correction for Random Ising
  Problems}}}, \bibinfo{journal}{Phys. Rev. A} \textbf{\bibinfo{volume}{91}},
  \bibinfo{pages}{042302} (\bibinfo{year}{2015}).

\bibitem[{com({\natexlab{d}})}]{comment:d-wave}
\bibinfo{note}{See {http://www.dwavesys.com}}.

\bibitem[{\citenamefont{Sidon}(1932)}]{sidon:32}
\bibinfo{author}{\bibfnamefont{S.}~\bibnamefont{Sidon}},
  \emph{\bibinfo{title}{{{Ein Satz {\"u}ber trigonometrische Polynome und seine
  Anwendung in der Theorie der Fourier-Reihen}}}},
  \bibinfo{journal}{Mathematische Annalen} \textbf{\bibinfo{volume}{106}},
  \bibinfo{pages}{536} (\bibinfo{year}{1932}).

\bibitem[{\citenamefont{Neuhaus}(2014)}]{neuhaus:14}
\bibinfo{author}{\bibfnamefont{T.}~\bibnamefont{Neuhaus}},
  \emph{\bibinfo{title}{{{Monte Carlo Search for Very Hard KSAT Realizations
  for Use in Quantum Annealing}}}} (\bibinfo{year}{2014}),
  \bibinfo{note}{(arXiv:cond-mat/1412.5361)}.

\bibitem[{com({\natexlab{e}})}]{comment:king}
\bibinfo{note}{Reference \cite{king:15} shows that if the instances are
  generated such that the coupling range is restricted, the results on DW2 are
  more favorable, as would be expected due to increased noise robustness.}

\bibitem[{\citenamefont{{Boixo}
  et~al.}(2014{\natexlab{b}})\citenamefont{{Boixo}, {Smelyanskiy}, {Shabani},
  {Isakov}, {Dykman}, {Denchev}, {Amin}, {Smirnov}, {Mohseni}, and
  {Neven}}}]{boixo:14a}
\bibinfo{author}{\bibfnamefont{S.}~\bibnamefont{{Boixo}}},
  \bibinfo{author}{\bibfnamefont{V.~N.} \bibnamefont{{Smelyanskiy}}},
  \bibinfo{author}{\bibfnamefont{A.}~\bibnamefont{{Shabani}}},
  \bibinfo{author}{\bibfnamefont{S.~V.} \bibnamefont{{Isakov}}},
  \bibinfo{author}{\bibfnamefont{M.}~\bibnamefont{{Dykman}}},
  \bibinfo{author}{\bibfnamefont{V.~S.} \bibnamefont{{Denchev}}},
  \bibinfo{author}{\bibfnamefont{M.}~\bibnamefont{{Amin}}},
  \bibinfo{author}{\bibfnamefont{A.}~\bibnamefont{{Smirnov}}},
  \bibinfo{author}{\bibfnamefont{M.}~\bibnamefont{{Mohseni}}},
  \bibnamefont{and} \bibinfo{author}{\bibfnamefont{H.}~\bibnamefont{{Neven}}},
  \emph{\bibinfo{title}{{{Computational Role of Collective Tunneling in a
  Quantum Annealer}}}} (\bibinfo{year}{2014}{\natexlab{b}}),
  \bibinfo{note}{arXiv:1411.4036}.

\bibitem[{\citenamefont{Hukushima and Nemoto}(1996)}]{hukushima:96}
\bibinfo{author}{\bibfnamefont{K.}~\bibnamefont{Hukushima}} \bibnamefont{and}
  \bibinfo{author}{\bibfnamefont{K.}~\bibnamefont{Nemoto}},
  \emph{\bibinfo{title}{Exchange {M}onte {C}arlo method and application to spin
  glass simulations}}, \bibinfo{journal}{J. Phys. Soc. Jpn.}
  \textbf{\bibinfo{volume}{65}}, \bibinfo{pages}{1604} (\bibinfo{year}{1996}).

\bibitem[{\citenamefont{Geyer}(1991)}]{geyer:91}
\bibinfo{author}{\bibfnamefont{C.}~\bibnamefont{Geyer}}, in
  \emph{\bibinfo{booktitle}{23rd Symposium on the Interface}}, edited by
  \bibinfo{editor}{\bibfnamefont{E.~M.} \bibnamefont{Keramidas}}
  (\bibinfo{publisher}{Interface Foundation}, \bibinfo{address}{Fairfax
  Station, VA}, \bibinfo{year}{1991}), p. \bibinfo{pages}{156}.

\bibitem[{\citenamefont{Katzgraber et~al.}(2006)\citenamefont{Katzgraber,
  Trebst, Huse, and Troyer}}]{katzgraber:06a}
\bibinfo{author}{\bibfnamefont{H.~G.} \bibnamefont{Katzgraber}},
  \bibinfo{author}{\bibfnamefont{S.}~\bibnamefont{Trebst}},
  \bibinfo{author}{\bibfnamefont{D.~A.} \bibnamefont{Huse}}, \bibnamefont{and}
  \bibinfo{author}{\bibfnamefont{M.}~\bibnamefont{Troyer}},
  \emph{\bibinfo{title}{{{Feedback-optimized parallel tempering Monte
  Carlo}}}}, \bibinfo{journal}{J. Stat. Mech.}
  \textbf{\bibinfo{volume}{\normalfont{P03018}}} (\bibinfo{year}{2006}).

\bibitem[{\citenamefont{Edwards and Anderson}(1975)}]{edwards:75}
\bibinfo{author}{\bibfnamefont{S.~F.} \bibnamefont{Edwards}} \bibnamefont{and}
  \bibinfo{author}{\bibfnamefont{P.~W.} \bibnamefont{Anderson}},
  \emph{\bibinfo{title}{Theory of spin glasses}}, \bibinfo{journal}{J. Phys. F:
  Met. Phys.} \textbf{\bibinfo{volume}{5}}, \bibinfo{pages}{965}
  (\bibinfo{year}{1975}).

\bibitem[{\citenamefont{Parisi}(1983)}]{parisi:83}
\bibinfo{author}{\bibfnamefont{G.}~\bibnamefont{Parisi}},
  \emph{\bibinfo{title}{Order parameter for spin-glasses}},
  \bibinfo{journal}{Phys. Rev. Lett.} \textbf{\bibinfo{volume}{50}},
  \bibinfo{pages}{1946} (\bibinfo{year}{1983}).

\bibitem[{com({\natexlab{f}})}]{comment:paths}
\bibinfo{note}{Clearly, phase space is a high-dimensional space. As such, it is
  very likely that there is always some path that will connect two points, thus
  avoiding a barrier. However, in this case, we refer to relatively
  straightforward paths, whatever that might mean in the frustrating world of
  spin glasses.}

\bibitem[{\citenamefont{Moreno et~al.}(2003)\citenamefont{Moreno, Katzgraber,
  and Hartmann}}]{moreno:03}
\bibinfo{author}{\bibfnamefont{J.~J.} \bibnamefont{Moreno}},
  \bibinfo{author}{\bibfnamefont{H.~G.} \bibnamefont{Katzgraber}},
  \bibnamefont{and} \bibinfo{author}{\bibfnamefont{A.~K.}
  \bibnamefont{Hartmann}}, \emph{\bibinfo{title}{Finding low-temperature states
  with parallel tempering, simulated annealing and simple {M}onte {C}arlo}},
  \bibinfo{journal}{Int. J. Mod. Phys. C} \textbf{\bibinfo{volume}{14}},
  \bibinfo{pages}{285} (\bibinfo{year}{2003}).

\bibitem[{\citenamefont{Katzgraber and Young}(2003)}]{katzgraber:03f}
\bibinfo{author}{\bibfnamefont{H.~G.} \bibnamefont{Katzgraber}}
  \bibnamefont{and} \bibinfo{author}{\bibfnamefont{A.~P.} \bibnamefont{Young}},
  \emph{\bibinfo{title}{Geometry of large-scale low-energy excitations in the
  one-dimensional {I}sing spin glass with power-law interactions}},
  \bibinfo{journal}{Phys. Rev. B} \textbf{\bibinfo{volume}{68}},
  \bibinfo{pages}{224408} (\bibinfo{year}{2003}).

\bibitem[{com({\natexlab{g}})}]{comment:s7}
\bibinfo{note}{In the meantime, we have discovered that for the vanilla Chimera
  graph the optimal Sidon set is U$_{5,6,7}$ with random interactions drawn
  from the set $\{\pm 5, \pm 6, \pm 7\}$. This set produces many instances with
  a low degeneracy and is also rather robust to the intrinsic noise of the DW2
  device because the classical energy gap $\Delta E = 2/7$ is considerably
  larger than the noise of the device. We report on these results in a
  subsequent publication \cite{zhu:15a}.}

\bibitem[{\citenamefont{Smolin and Smith}(2013)}]{smolin:13}
\bibinfo{author}{\bibfnamefont{J.~A.} \bibnamefont{Smolin}} \bibnamefont{and}
  \bibinfo{author}{\bibfnamefont{G.}~\bibnamefont{Smith}},
  \emph{\bibinfo{title}{Classical signature of quantum annealing}},
  \bibinfo{journal}{arXiv preprint arXiv:1305.4904}  (\bibinfo{year}{2013}).

\bibitem[{\citenamefont{Shin et~al.}(2014)\citenamefont{Shin, Smith, Smolin,
  and Vazirani}}]{shin:14}
\bibinfo{author}{\bibfnamefont{S.~W.} \bibnamefont{Shin}},
  \bibinfo{author}{\bibfnamefont{G.}~\bibnamefont{Smith}},
  \bibinfo{author}{\bibfnamefont{J.~A.} \bibnamefont{Smolin}},
  \bibnamefont{and} \bibinfo{author}{\bibfnamefont{U.}~\bibnamefont{Vazirani}},
  \emph{\bibinfo{title}{{{How ``Quantum'' is the D-Wave Machine?}}}}
  (\bibinfo{year}{2014}), \bibinfo{note}{(arXiv:1401.7087)}.

\bibitem[{\citenamefont{Bishop}(2006)}]{bishop:06}
\bibinfo{author}{\bibfnamefont{C.}~\bibnamefont{Bishop}},
  \emph{\bibinfo{title}{Pattern {R}ecognition and {M}achine {L}earning}}
  (\bibinfo{publisher}{Springer-Verlag}, \bibinfo{address}{New York},
  \bibinfo{year}{2006}).

\bibitem[{\citenamefont{Gelman et~al.}(2003)\citenamefont{Gelman, Carlin,
  Stern, and Rubin}}]{gelman:03}
\bibinfo{author}{\bibfnamefont{A.}~\bibnamefont{Gelman}},
  \bibinfo{author}{\bibfnamefont{J.~B.} \bibnamefont{Carlin}},
  \bibinfo{author}{\bibfnamefont{H.~L.} \bibnamefont{Stern}}, \bibnamefont{and}
  \bibinfo{author}{\bibfnamefont{D.~B.} \bibnamefont{Rubin}},
  \emph{\bibinfo{title}{Bayesian Data Analysis}} (\bibinfo{publisher}{Chapman
  and Hall/CRC}, \bibinfo{address}{London}, \bibinfo{year}{2003}).

\bibitem[{\citenamefont{Houdayer and
  Martin}(1999{\natexlab{a}})}]{houdayer:99a}
\bibinfo{author}{\bibfnamefont{J.}~\bibnamefont{Houdayer}} \bibnamefont{and}
  \bibinfo{author}{\bibfnamefont{O.~C.} \bibnamefont{Martin}},
  \emph{\bibinfo{title}{Renormalization for discrete optimization}},
  \bibinfo{journal}{Phys. Rev. Lett.} \textbf{\bibinfo{volume}{83}},
  \bibinfo{pages}{1030} (\bibinfo{year}{1999}{\natexlab{a}}).

\bibitem[{\citenamefont{Thomas et~al.}(2008)\citenamefont{Thomas, White, and
  Middleton}}]{thomas:08}
\bibinfo{author}{\bibfnamefont{C.~K.} \bibnamefont{Thomas}},
  \bibinfo{author}{\bibfnamefont{O.~L.} \bibnamefont{White}}, \bibnamefont{and}
  \bibinfo{author}{\bibfnamefont{A.~A.} \bibnamefont{Middleton}},
  \emph{\bibinfo{title}{{{Persistence and memory in patchwork dynamics for
  glassy models}}}}, \bibinfo{journal}{Phys. Rev. B}
  \textbf{\bibinfo{volume}{77}}, \bibinfo{pages}{092415}
  (\bibinfo{year}{2008}).

\bibitem[{\citenamefont{{Zintchenko} et~al.}(2015)\citenamefont{{Zintchenko},
  {Hastings}, and {Troyer}}}]{zintchenko:15}
\bibinfo{author}{\bibfnamefont{I.}~\bibnamefont{{Zintchenko}}},
  \bibinfo{author}{\bibfnamefont{M.~B.} \bibnamefont{{Hastings}}},
  \bibnamefont{and} \bibinfo{author}{\bibfnamefont{M.}~\bibnamefont{{Troyer}}},
  \emph{\bibinfo{title}{{From local to global ground states in Ising spin
  glasses}}}, \bibinfo{journal}{Phys. Rev. B} \textbf{\bibinfo{volume}{91}},
  \bibinfo{pages}{024201} (\bibinfo{year}{2015}).

\bibitem[{\citenamefont{{Belletti} et~al.}(2008)\citenamefont{{Belletti},
  {Cotallo}, {Cruz}, {Fern{\'a}ndez}, {Gordillo}, {Maiorano}, {Mantovani},
  {Marinari}, {Mart{\'{\i}}n-Mayor}, {Mu{\~n}oz-Sudupe} et~al.}}]{belletti:08}
\bibinfo{author}{\bibfnamefont{F.}~\bibnamefont{{Belletti}}},
  \bibinfo{author}{\bibfnamefont{M.}~\bibnamefont{{Cotallo}}},
  \bibinfo{author}{\bibfnamefont{A.}~\bibnamefont{{Cruz}}},
  \bibinfo{author}{\bibfnamefont{L.~A.} \bibnamefont{{Fern{\'a}ndez}}},
  \bibinfo{author}{\bibfnamefont{A.}~\bibnamefont{{Gordillo}}},
  \bibinfo{author}{\bibfnamefont{A.}~\bibnamefont{{Maiorano}}},
  \bibinfo{author}{\bibfnamefont{F.}~\bibnamefont{{Mantovani}}},
  \bibinfo{author}{\bibfnamefont{E.}~\bibnamefont{{Marinari}}},
  \bibinfo{author}{\bibfnamefont{V.}~\bibnamefont{{Mart{\'{\i}}n-Mayor}}},
  \bibinfo{author}{\bibfnamefont{A.}~\bibnamefont{{Mu{\~n}oz-Sudupe}}},
  \bibnamefont{et~al.}, \emph{\bibinfo{title}{{{Simulating spin systems on
  IANUS, an FPGA-based computer}}}}, \bibinfo{journal}{Comp. Phys. Comm.}
  \textbf{\bibinfo{volume}{178}}, \bibinfo{pages}{208} (\bibinfo{year}{2008}).

\bibitem[{\citenamefont{{McKay} et~al.}(1982)\citenamefont{{McKay}, {Berker},
  and {Kirkpatrick}}}]{mckay:82}
\bibinfo{author}{\bibfnamefont{S.~R.} \bibnamefont{{McKay}}},
  \bibinfo{author}{\bibfnamefont{A.~N.} \bibnamefont{{Berker}}},
  \bibnamefont{and}
  \bibinfo{author}{\bibfnamefont{S.}~\bibnamefont{{Kirkpatrick}}},
  \emph{\bibinfo{title}{{{Spin-Glass Behavior in Frustrated Ising Models with
  Chaotic Renormalization-Group Trajectories}}}}, \bibinfo{journal}{Phys. Rev.
  Lett.} \textbf{\bibinfo{volume}{48}}, \bibinfo{pages}{767}
  (\bibinfo{year}{1982}).

\bibitem[{\citenamefont{{Parisi}}(1984)}]{parisi:84}
\bibinfo{author}{\bibfnamefont{G.}~\bibnamefont{{Parisi}}},
  \emph{\bibinfo{title}{{Spin glasses and replicas}}},
  \bibinfo{journal}{Physica A} \textbf{\bibinfo{volume}{124}},
  \bibinfo{pages}{523} (\bibinfo{year}{1984}).

\bibitem[{\citenamefont{Fisher and Huse}(1986)}]{fisher:86}
\bibinfo{author}{\bibfnamefont{D.~S.} \bibnamefont{Fisher}} \bibnamefont{and}
  \bibinfo{author}{\bibfnamefont{D.~A.} \bibnamefont{Huse}},
  \emph{\bibinfo{title}{Ordered phase of short-range {I}sing spin-glasses}},
  \bibinfo{journal}{Phys. Rev. Lett.} \textbf{\bibinfo{volume}{56}},
  \bibinfo{pages}{1601} (\bibinfo{year}{1986}).

\bibitem[{\citenamefont{Bray and Moore}(1987)}]{bray:87}
\bibinfo{author}{\bibfnamefont{A.~J.} \bibnamefont{Bray}} \bibnamefont{and}
  \bibinfo{author}{\bibfnamefont{M.~A.} \bibnamefont{Moore}},
  \emph{\bibinfo{title}{{Chaotic Nature of the Spin-Glass Phase}}},
  \bibinfo{journal}{Phys. Rev. Lett.} \textbf{\bibinfo{volume}{58}},
  \bibinfo{pages}{57} (\bibinfo{year}{1987}).

\bibitem[{\citenamefont{Kondor}(1989)}]{kondor:89}
\bibinfo{author}{\bibfnamefont{I.}~\bibnamefont{Kondor}},
  \emph{\bibinfo{title}{{{On chaos in spin glasses}}}}, \bibinfo{journal}{J.
  Phys. A} \textbf{\bibinfo{volume}{22}}, \bibinfo{pages}{L163}
  (\bibinfo{year}{1989}).

\bibitem[{\citenamefont{{Ney-Nifle} and {Young}}(1997)}]{neynifle:97}
\bibinfo{author}{\bibfnamefont{M.}~\bibnamefont{{Ney-Nifle}}} \bibnamefont{and}
  \bibinfo{author}{\bibfnamefont{A.~P.} \bibnamefont{{Young}}},
  \emph{\bibinfo{title}{{{Chaos in a two-dimensional Ising spin glass}}}},
  \bibinfo{journal}{J. Phys. A} \textbf{\bibinfo{volume}{30}},
  \bibinfo{pages}{5311} (\bibinfo{year}{1997}).

\bibitem[{\citenamefont{{Ney-Nifle}}(1998)}]{neynifle:98}
\bibinfo{author}{\bibfnamefont{M.}~\bibnamefont{{Ney-Nifle}}},
  \emph{\bibinfo{title}{{Chaos and universality in a four-dimensional spin
  glass}}}, \bibinfo{journal}{Phys. Rev. B} \textbf{\bibinfo{volume}{57}},
  \bibinfo{pages}{492} (\bibinfo{year}{1998}).

\bibitem[{\citenamefont{{Billoire} and {Marinari}}(2000)}]{billoire:00}
\bibinfo{author}{\bibfnamefont{A.}~\bibnamefont{{Billoire}}} \bibnamefont{and}
  \bibinfo{author}{\bibfnamefont{E.}~\bibnamefont{{Marinari}}},
  \emph{\bibinfo{title}{{{Evidence against temperature chaos in mean-field and
  realistic spin glasses}}}}, \bibinfo{journal}{J. Phys. A}
  \textbf{\bibinfo{volume}{33}}, \bibinfo{pages}{L265} (\bibinfo{year}{2000}).

\bibitem[{\citenamefont{{Billoire} and {Marinari}}(2002)}]{billoire:02}
\bibinfo{author}{\bibfnamefont{A.}~\bibnamefont{{Billoire}}} \bibnamefont{and}
  \bibinfo{author}{\bibfnamefont{E.}~\bibnamefont{{Marinari}}},
  \emph{\bibinfo{title}{{{Overlap among states at different temperatures in the
  SK model}}}}, \bibinfo{journal}{Europhys. Lett.}
  \textbf{\bibinfo{volume}{60}}, \bibinfo{pages}{775} (\bibinfo{year}{2002}).

\bibitem[{\citenamefont{{Sasaki} et~al.}(2005)\citenamefont{{Sasaki},
  {Hukushima}, {Yoshino}, and {Takayama}}}]{sasaki:05}
\bibinfo{author}{\bibfnamefont{M.}~\bibnamefont{{Sasaki}}},
  \bibinfo{author}{\bibfnamefont{K.}~\bibnamefont{{Hukushima}}},
  \bibinfo{author}{\bibfnamefont{H.}~\bibnamefont{{Yoshino}}},
  \bibnamefont{and}
  \bibinfo{author}{\bibfnamefont{H.}~\bibnamefont{{Takayama}}},
  \emph{\bibinfo{title}{{{Temperature Chaos and Bond Chaos in Edwards-Anderson
  Ising Spin Glasses: Domain-Wall Free-Energy Measurements}}}},
  \bibinfo{journal}{Phys. Rev. Lett.} \textbf{\bibinfo{volume}{95}},
  \bibinfo{pages}{267203} (\bibinfo{year}{2005}).

\bibitem[{\citenamefont{Katzgraber and Krzakala}(2007)}]{katzgraber:07}
\bibinfo{author}{\bibfnamefont{H.~G.} \bibnamefont{Katzgraber}}
  \bibnamefont{and} \bibinfo{author}{\bibfnamefont{F.}~\bibnamefont{Krzakala}},
  \emph{\bibinfo{title}{{{Temperature and Disorder Chaos in Three-Dimensional
  Ising Spin Glasses}}}}, \bibinfo{journal}{Phys. Rev. Lett.}
  \textbf{\bibinfo{volume}{98}}, \bibinfo{pages}{017201}
  (\bibinfo{year}{2007}).

\bibitem[{com({\natexlab{h}})}]{comment:chicken}
\bibinfo{note}{In fact, a recent study similar to our preliminary results (see,
  e.g., {https://youtu.be/C8fSpHW9XHk}) attempts to design harder benchmark
  instances by exploiting the chaotic effects in spin-glass systems
  \cite{martin-mayor:15}.}

\bibitem[{\citenamefont{Parisi}(1980)}]{parisi:80}
\bibinfo{author}{\bibfnamefont{G.}~\bibnamefont{Parisi}},
  \emph{\bibinfo{title}{The order parameter for spin glasses: a function on the
  interval $0$--$1$}}, \bibinfo{journal}{J. Phys. A}
  \textbf{\bibinfo{volume}{13}}, \bibinfo{pages}{1101} (\bibinfo{year}{1980}).

\bibitem[{\citenamefont{de~Almeida and Thouless}(1978)}]{almeida:78}
\bibinfo{author}{\bibfnamefont{J.~R.~L.} \bibnamefont{de~Almeida}}
  \bibnamefont{and} \bibinfo{author}{\bibfnamefont{D.~J.}
  \bibnamefont{Thouless}}, \emph{\bibinfo{title}{Stability of the
  {S}herrington-{K}irkpatrick solution of a spin glass model}},
  \bibinfo{journal}{J. Phys. A} \textbf{\bibinfo{volume}{11}},
  \bibinfo{pages}{983} (\bibinfo{year}{1978}).

\bibitem[{\citenamefont{Bhatt and Young}(1985)}]{bhatt:85}
\bibinfo{author}{\bibfnamefont{R.~N.} \bibnamefont{Bhatt}} \bibnamefont{and}
  \bibinfo{author}{\bibfnamefont{A.~P.} \bibnamefont{Young}},
  \emph{\bibinfo{title}{Search for a transition in the three-dimensional $\pm
  {J}$ {I}sing spin-glass}}, \bibinfo{journal}{Phys. Rev. Lett.}
  \textbf{\bibinfo{volume}{54}}, \bibinfo{pages}{924} (\bibinfo{year}{1985}).

\bibitem[{\citenamefont{Billoire and Coluzzi}(2003)}]{billoire:03b}
\bibinfo{author}{\bibfnamefont{A.}~\bibnamefont{Billoire}} \bibnamefont{and}
  \bibinfo{author}{\bibfnamefont{B.}~\bibnamefont{Coluzzi}},
  \emph{\bibinfo{title}{{Numerical study of the {S}herrington-{K}irkpatrick
  model in a magnetic field}}}, \bibinfo{journal}{Phys. Rev. E}
  \textbf{\bibinfo{volume}{68}}, \bibinfo{pages}{026131}
  (\bibinfo{year}{2003}).

\bibitem[{\citenamefont{Barrat and Berthier}(2001)}]{barrat:01}
\bibinfo{author}{\bibfnamefont{A.}~\bibnamefont{Barrat}} \bibnamefont{and}
  \bibinfo{author}{\bibfnamefont{L.}~\bibnamefont{Berthier}},
  \emph{\bibinfo{title}{Real-space application of the mean-field description of
  spin-glass dynamics}}, \bibinfo{journal}{Phys. Rev. Lett.}
  \textbf{\bibinfo{volume}{87}}, \bibinfo{pages}{087204}
  (\bibinfo{year}{2001}).

\bibitem[{\citenamefont{Takayama and Hukushima}(2004)}]{takayama:04}
\bibinfo{author}{\bibfnamefont{H.}~\bibnamefont{Takayama}} \bibnamefont{and}
  \bibinfo{author}{\bibfnamefont{K.}~\bibnamefont{Hukushima}},
  \emph{\bibinfo{title}{{Field-shift aging protocol on the 3D {I}sing
  spin-glass model: dynamical crossover between the spin-glass and paramagnetic
  states}}}, \bibinfo{journal}{J. Phys. Soc. Jpn.}
  \textbf{\bibinfo{volume}{73}}, \bibinfo{pages}{2077} (\bibinfo{year}{2004}).

\bibitem[{\citenamefont{Houdayer and Martin}(1999{\natexlab{b}})}]{houdayer:99}
\bibinfo{author}{\bibfnamefont{J.}~\bibnamefont{Houdayer}} \bibnamefont{and}
  \bibinfo{author}{\bibfnamefont{O.~C.} \bibnamefont{Martin}},
  \emph{\bibinfo{title}{{I}sing spin glasses in a magnetic field}},
  \bibinfo{journal}{Phys. Rev. Lett.} \textbf{\bibinfo{volume}{82}},
  \bibinfo{pages}{4934} (\bibinfo{year}{1999}{\natexlab{b}}).

\bibitem[{\citenamefont{Krzakala et~al.}(2001)\citenamefont{Krzakala, Houdayer,
  Marinari, Martin, and Parisi}}]{krzakala:01}
\bibinfo{author}{\bibfnamefont{F.}~\bibnamefont{Krzakala}},
  \bibinfo{author}{\bibfnamefont{J.}~\bibnamefont{Houdayer}},
  \bibinfo{author}{\bibfnamefont{E.}~\bibnamefont{Marinari}},
  \bibinfo{author}{\bibfnamefont{O.~C.} \bibnamefont{Martin}},
  \bibnamefont{and} \bibinfo{author}{\bibfnamefont{G.}~\bibnamefont{Parisi}},
  \emph{\bibinfo{title}{{Zero-temperature responses of a 3D spin glass in a
  field}}}, \bibinfo{journal}{Phys. Rev. Lett.} \textbf{\bibinfo{volume}{87}},
  \bibinfo{pages}{197204} (\bibinfo{year}{2001}).

\bibitem[{\citenamefont{{Leuzzi} et~al.}(2009)\citenamefont{{Leuzzi}, {Parisi},
  {Ricci-Tersenghi}, and {Ruiz-Lorenzo}}}]{leuzzi:09}
\bibinfo{author}{\bibfnamefont{L.}~\bibnamefont{{Leuzzi}}},
  \bibinfo{author}{\bibfnamefont{G.}~\bibnamefont{{Parisi}}},
  \bibinfo{author}{\bibfnamefont{F.}~\bibnamefont{{Ricci-Tersenghi}}},
  \bibnamefont{and} \bibinfo{author}{\bibfnamefont{J.~J.}
  \bibnamefont{{Ruiz-Lorenzo}}}, \emph{\bibinfo{title}{{{Ising Spin-Glass
  Transition in a Magnetic Field Outside the Limit of Validity of Mean-Field
  Theory}}}}, \bibinfo{journal}{Phys. Rev. Lett.}
  \textbf{\bibinfo{volume}{103}}, \bibinfo{pages}{267201}
  (\bibinfo{year}{2009}).

\bibitem[{\citenamefont{{Ba{\~n}os} et~al.}(2012)\citenamefont{{Ba{\~n}os},
  {Cruz}, {Fernandez}, {Gil-Narvion}, {Gordillo-Guerrero}, {Guidetti},
  {I{\~n}iguez}, {Maiorano}, {Marinari}, {Martin-Mayor} et~al.}}]{banos:12}
\bibinfo{author}{\bibfnamefont{R.~A.} \bibnamefont{{Ba{\~n}os}}},
  \bibinfo{author}{\bibfnamefont{A.}~\bibnamefont{{Cruz}}},
  \bibinfo{author}{\bibfnamefont{L.~A.} \bibnamefont{{Fernandez}}},
  \bibinfo{author}{\bibfnamefont{J.~M.} \bibnamefont{{Gil-Narvion}}},
  \bibinfo{author}{\bibfnamefont{A.}~\bibnamefont{{Gordillo-Guerrero}}},
  \bibinfo{author}{\bibfnamefont{M.}~\bibnamefont{{Guidetti}}},
  \bibinfo{author}{\bibfnamefont{D.}~\bibnamefont{{I{\~n}iguez}}},
  \bibinfo{author}{\bibfnamefont{A.}~\bibnamefont{{Maiorano}}},
  \bibinfo{author}{\bibfnamefont{E.}~\bibnamefont{{Marinari}}},
  \bibinfo{author}{\bibfnamefont{V.}~\bibnamefont{{Martin-Mayor}}},
  \bibnamefont{et~al.}, \emph{\bibinfo{title}{{{Thermodynamic glass transition
  in a spin glass without time-reversal symmetry}}}}, \bibinfo{journal}{Proc.
  Natl. Acad. Sci. U.S.A.} \textbf{\bibinfo{volume}{109}},
  \bibinfo{pages}{6452} (\bibinfo{year}{2012}).

\bibitem[{\citenamefont{{Baity-Jesi} et~al.}(2014)\citenamefont{{Baity-Jesi},
  {Alvarez Ba{\~n}os}, {Cruz}, {Fernandez}, {Gil-Narvion}, {Gordillo-Guerrero},
  {I{\~n}iguez}, {Maiorano}, {Mantovani}, {Marinari} et~al.}}]{baity:14}
\bibinfo{author}{\bibfnamefont{M.}~\bibnamefont{{Baity-Jesi}}},
  \bibinfo{author}{\bibfnamefont{R.}~\bibnamefont{{Alvarez Ba{\~n}os}}},
  \bibinfo{author}{\bibfnamefont{A.}~\bibnamefont{{Cruz}}},
  \bibinfo{author}{\bibfnamefont{L.~A.} \bibnamefont{{Fernandez}}},
  \bibinfo{author}{\bibfnamefont{J.~M.} \bibnamefont{{Gil-Narvion}}},
  \bibinfo{author}{\bibnamefont{{Gordillo-Guerrero}}},
  \bibinfo{author}{\bibfnamefont{D.}~\bibnamefont{{I{\~n}iguez}}},
  \bibinfo{author}{\bibfnamefont{A.}~\bibnamefont{{Maiorano}}},
  \bibinfo{author}{\bibfnamefont{F.}~\bibnamefont{{Mantovani}}},
  \bibinfo{author}{\bibfnamefont{E.}~\bibnamefont{{Marinari}}},
  \bibnamefont{et~al.}, \emph{\bibinfo{title}{{{Dynamical transition in the
  $D=3$ Edwards-Anderson spin glass in an external magnetic field}}}},
  \bibinfo{journal}{Phys. Rev. E} \textbf{\bibinfo{volume}{89}},
  \bibinfo{pages}{032140} (\bibinfo{year}{2014}).

\bibitem[{\citenamefont{Young and Katzgraber}(2004)}]{young:04}
\bibinfo{author}{\bibfnamefont{A.~P.} \bibnamefont{Young}} \bibnamefont{and}
  \bibinfo{author}{\bibfnamefont{H.~G.} \bibnamefont{Katzgraber}},
  \emph{\bibinfo{title}{{Absence of an Almeida-Thouless line in
  Three-Dimensional Spin Glasses}}}, \bibinfo{journal}{Phys. Rev. Lett.}
  \textbf{\bibinfo{volume}{93}}, \bibinfo{pages}{207203}
  (\bibinfo{year}{2004}).

\bibitem[{\citenamefont{Katzgraber and {Young}}(2005)}]{katzgraber:05c}
\bibinfo{author}{\bibfnamefont{H.~G.} \bibnamefont{Katzgraber}}
  \bibnamefont{and} \bibinfo{author}{\bibfnamefont{A.~P.}
  \bibnamefont{{Young}}}, \emph{\bibinfo{title}{{{Probing the Almeida-Thouless
  line away from the mean-field model}}}}, \bibinfo{journal}{Phys. Rev. B}
  \textbf{\bibinfo{volume}{72}}, \bibinfo{pages}{184416}
  (\bibinfo{year}{2005}).

\bibitem[{\citenamefont{Katzgraber et~al.}(2009)\citenamefont{Katzgraber,
  Larson, and Young}}]{katzgraber:09b}
\bibinfo{author}{\bibfnamefont{H.~G.} \bibnamefont{Katzgraber}},
  \bibinfo{author}{\bibfnamefont{D.}~\bibnamefont{Larson}}, \bibnamefont{and}
  \bibinfo{author}{\bibfnamefont{A.~P.} \bibnamefont{Young}},
  \emph{\bibinfo{title}{Study of the de {A}lmeida-{T}houless line using
  power-law diluted one-dimensional {I}sing spin glasses}},
  \bibinfo{journal}{Phys. Rev. Lett.} \textbf{\bibinfo{volume}{102}},
  \bibinfo{pages}{177205} (\bibinfo{year}{2009}).

\bibitem[{\citenamefont{Larson et~al.}(2013)\citenamefont{Larson, Katzgraber,
  Moore, and Young}}]{larson:13}
\bibinfo{author}{\bibfnamefont{D.}~\bibnamefont{Larson}},
  \bibinfo{author}{\bibfnamefont{H.~G.} \bibnamefont{Katzgraber}},
  \bibinfo{author}{\bibfnamefont{M.~A.} \bibnamefont{Moore}}, \bibnamefont{and}
  \bibinfo{author}{\bibfnamefont{A.~P.} \bibnamefont{Young}},
  \emph{\bibinfo{title}{{Spin glasses in a field: Three and four dimensions as
  seen from one space dimension}}}, \bibinfo{journal}{Phys. Rev. B}
  \textbf{\bibinfo{volume}{87}}, \bibinfo{pages}{024414}
  (\bibinfo{year}{2013}).

\bibitem[{\citenamefont{{Perdomo-Ortiz}
  et~al.}(2015{\natexlab{a}})\citenamefont{{Perdomo-Ortiz}, {O'Gorman},
  {Fluegemann}, {Biswas}, and {Smelyanskiy}}}]{perdomo:15}
\bibinfo{author}{\bibfnamefont{A.}~\bibnamefont{{Perdomo-Ortiz}}},
  \bibinfo{author}{\bibfnamefont{B.}~\bibnamefont{{O'Gorman}}},
  \bibinfo{author}{\bibfnamefont{J.}~\bibnamefont{{Fluegemann}}},
  \bibinfo{author}{\bibfnamefont{R.}~\bibnamefont{{Biswas}}}, \bibnamefont{and}
  \bibinfo{author}{\bibfnamefont{V.~N.} \bibnamefont{{Smelyanskiy}}},
  \emph{\bibinfo{title}{{Determination and correction of persistent biases in
  quantum annealers}}} (\bibinfo{year}{2015}{\natexlab{a}}),
  \bibinfo{note}{(arXiv:quant-phys/1503.05679)}.

\bibitem[{\citenamefont{{Perdomo-Ortiz}
  et~al.}(2015{\natexlab{b}})\citenamefont{{Perdomo-Ortiz}, {Fluegemann},
  {Biswas}, and {Smelyanskiy}}}]{perdomo:15a}
\bibinfo{author}{\bibfnamefont{A.}~\bibnamefont{{Perdomo-Ortiz}}},
  \bibinfo{author}{\bibfnamefont{J.}~\bibnamefont{{Fluegemann}}},
  \bibinfo{author}{\bibfnamefont{R.}~\bibnamefont{{Biswas}}}, \bibnamefont{and}
  \bibinfo{author}{\bibfnamefont{V.~N.} \bibnamefont{{Smelyanskiy}}},
  \emph{\bibinfo{title}{{{A Performance Estimator for Quantum Annealers: Gauge
  selection and Parameter Setting}}}} (\bibinfo{year}{2015}{\natexlab{b}}),
  \bibinfo{note}{(arXiv:quant-phys/1503.01083)}.

\bibitem[{\citenamefont{Smith and Smolin}(2013)}]{smith:13}
\bibinfo{author}{\bibfnamefont{G.}~\bibnamefont{Smith}} \bibnamefont{and}
  \bibinfo{author}{\bibfnamefont{J.}~\bibnamefont{Smolin}},
  \emph{\bibinfo{title}{{{Putting ``Quantumness'' to the Test}}}},
  \bibinfo{journal}{Physics} \textbf{\bibinfo{volume}{6}}, \bibinfo{pages}{105}
  (\bibinfo{year}{2013}).

\bibitem[{\citenamefont{{Lanting} et~al.}(2014)\citenamefont{{Lanting},
  {Przybysz}, {Smirnov}, {Spedalieri}, {Amin}, {Berkley}, {Harris}, {Altomare},
  {Boixo}, {Bunyk} et~al.}}]{lanting:14}
\bibinfo{author}{\bibfnamefont{T.}~\bibnamefont{{Lanting}}},
  \bibinfo{author}{\bibfnamefont{A.~J.} \bibnamefont{{Przybysz}}},
  \bibinfo{author}{\bibfnamefont{A.~Y.} \bibnamefont{{Smirnov}}},
  \bibinfo{author}{\bibfnamefont{F.~M.} \bibnamefont{{Spedalieri}}},
  \bibinfo{author}{\bibfnamefont{M.~H.} \bibnamefont{{Amin}}},
  \bibinfo{author}{\bibfnamefont{A.~J.} \bibnamefont{{Berkley}}},
  \bibinfo{author}{\bibfnamefont{R.}~\bibnamefont{{Harris}}},
  \bibinfo{author}{\bibfnamefont{F.}~\bibnamefont{{Altomare}}},
  \bibinfo{author}{\bibfnamefont{S.}~\bibnamefont{{Boixo}}},
  \bibinfo{author}{\bibfnamefont{P.}~\bibnamefont{{Bunyk}}},
  \bibnamefont{et~al.}, \emph{\bibinfo{title}{Entanglement in a quantum
  annealing processor}}, \bibinfo{journal}{Phys. Rev. X}
  \textbf{\bibinfo{volume}{4}}, \bibinfo{pages}{021041} (\bibinfo{year}{2014}).

\bibitem[{\citenamefont{Albash et~al.}(2015{\natexlab{a}})\citenamefont{Albash,
  Vinci, Mishra, Warburton, and Lidar}}]{albash:15}
\bibinfo{author}{\bibfnamefont{T.}~\bibnamefont{Albash}},
  \bibinfo{author}{\bibfnamefont{W.}~\bibnamefont{Vinci}},
  \bibinfo{author}{\bibfnamefont{A.}~\bibnamefont{Mishra}},
  \bibinfo{author}{\bibfnamefont{P.~A.} \bibnamefont{Warburton}},
  \bibnamefont{and} \bibinfo{author}{\bibfnamefont{D.~A.} \bibnamefont{Lidar}},
  \emph{\bibinfo{title}{{{Consistency Tests of Classical and Quantum Models for
  a Quantum Device}}}}, \bibinfo{journal}{Phys. Rev. A}
  \textbf{\bibinfo{volume}{91}}, \bibinfo{pages}{042314}
  (\bibinfo{year}{2015}{\natexlab{a}}).

\bibitem[{\citenamefont{Albash et~al.}(2015{\natexlab{b}})\citenamefont{Albash,
  R{\o}nnow, Troyer, and Lidar}}]{albash:15a}
\bibinfo{author}{\bibfnamefont{T.}~\bibnamefont{Albash}},
  \bibinfo{author}{\bibfnamefont{T.~F.} \bibnamefont{R{\o}nnow}},
  \bibinfo{author}{\bibfnamefont{M.}~\bibnamefont{Troyer}}, \bibnamefont{and}
  \bibinfo{author}{\bibfnamefont{D.~A.} \bibnamefont{Lidar}},
  \emph{\bibinfo{title}{{{Reexamining classical and quantum models for the
  D-Wave One processor}}}}, \bibinfo{journal}{Eur. Phys. J. Spec. Top.}
  \textbf{\bibinfo{volume}{224}}, \bibinfo{pages}{111}
  (\bibinfo{year}{2015}{\natexlab{b}}).

\bibitem[{\citenamefont{Kobe and Klotz}(1995)}]{kobe:95}
\bibinfo{author}{\bibfnamefont{S.}~\bibnamefont{Kobe}} \bibnamefont{and}
  \bibinfo{author}{\bibfnamefont{T.}~\bibnamefont{Klotz}},
  \emph{\bibinfo{title}{{Frustration: How it can be measured}}},
  \bibinfo{journal}{Phys. Rev. E} \textbf{\bibinfo{volume}{52}},
  \bibinfo{pages}{5660} (\bibinfo{year}{1995}).

\bibitem[{\citenamefont{Melchert~{\em et al.}}(2015)}]{melchert:15}
\bibinfo{author}{\bibfnamefont{O.}~\bibnamefont{Melchert~{\em et al.}}},
  \bibinfo{journal}{in preparation}  (\bibinfo{year}{2015}).

\bibitem[{\citenamefont{{Zhu} et~al.}(2015)\citenamefont{{Zhu}, {Ochoa}, and
  {Katzgraber}}}]{zhu:15b}
\bibinfo{author}{\bibfnamefont{Z.}~\bibnamefont{{Zhu}}},
  \bibinfo{author}{\bibfnamefont{A.~J.} \bibnamefont{{Ochoa}}},
  \bibnamefont{and} \bibinfo{author}{\bibfnamefont{H.~G.}
  \bibnamefont{{Katzgraber}}}, \emph{\bibinfo{title}{{{Efficient Cluster
  Algorithm for Spin Glasses in Any Space Dimension}}}},
  \bibinfo{journal}{Phys. Rev. Lett.} \textbf{\bibinfo{volume}{115}},
  \bibinfo{pages}{077201} (\bibinfo{year}{2015}).

\bibitem[{\citenamefont{Katzgraber}(2009)}]{katzgraber:09e}
\bibinfo{author}{\bibfnamefont{H.~G.} \bibnamefont{Katzgraber}},
  \emph{\bibinfo{title}{{{Introduction to Monte Carlo Methods}}}}
  (\bibinfo{year}{2009}), \bibinfo{note}{(arXiv:0905.1629)}.

\bibitem[{\citenamefont{Middleton}(2004)}]{middleton:04}
\bibinfo{author}{\bibfnamefont{A.~A.} \bibnamefont{Middleton}},
  \emph{\bibinfo{title}{Improved extremal optimization for the ising spin
  glass}}, \bibinfo{journal}{Phys. Rev. E} \textbf{\bibinfo{volume}{69}},
  \bibinfo{pages}{055701(R)} (\bibinfo{year}{2004}).

\bibitem[{\citenamefont{{Boettcher} and {Percus}}(2001)}]{boettcher:01}
\bibinfo{author}{\bibfnamefont{S.}~\bibnamefont{{Boettcher}}} \bibnamefont{and}
  \bibinfo{author}{\bibfnamefont{A.~G.} \bibnamefont{{Percus}}},
  \emph{\bibinfo{title}{{{Optimization with Extremal Dynamics}}}},
  \bibinfo{journal}{Phys. Rev. Lett.} \textbf{\bibinfo{volume}{86}},
  \bibinfo{pages}{5211} (\bibinfo{year}{2001}).

\bibitem[{\citenamefont{Weigel et~al.}(2015)\citenamefont{Weigel, Katzgraber,
  Machta, Hamze, Andrist, and {Octomore Collaboration}}}]{weigel:15}
\bibinfo{author}{\bibfnamefont{M.}~\bibnamefont{Weigel}},
  \bibinfo{author}{\bibfnamefont{H.~G.} \bibnamefont{Katzgraber}},
  \bibinfo{author}{\bibfnamefont{J.}~\bibnamefont{Machta}},
  \bibinfo{author}{\bibfnamefont{F.}~\bibnamefont{Hamze}},
  \bibinfo{author}{\bibfnamefont{R.~S.} \bibnamefont{Andrist}},
  \bibnamefont{and} \bibinfo{author}{\bibnamefont{{Octomore Collaboration}}},
  \emph{\bibinfo{title}{{Erratum: Glassy Chimeras could be blind to quantum
  speedup: Designing better benchmarks for quantum annealing machines [Phys.
  Rev. X 4, 021008 (2014)]}}}, \bibinfo{journal}{Phys. Rev. X}
  \textbf{\bibinfo{volume}{5}}, \bibinfo{pages}{019901} (\bibinfo{year}{2015}).

\bibitem[{\citenamefont{King}(2015)}]{king:15}
\bibinfo{author}{\bibfnamefont{A.~D.} \bibnamefont{King}},
  \emph{\bibinfo{title}{{{Performance of a quantum annealer on range-limited
  constraint satisfaction problems}}}} (\bibinfo{year}{2015}),
  \bibinfo{note}{arXiv:1502.02098}.

\bibitem[{\citenamefont{{Martin-Mayor} and {Hen}}(2015)}]{martin-mayor:15}
\bibinfo{author}{\bibfnamefont{V.}~\bibnamefont{{Martin-Mayor}}}
  \bibnamefont{and} \bibinfo{author}{\bibfnamefont{I.}~\bibnamefont{{Hen}}},
  \emph{\bibinfo{title}{{{Unraveling Quantum Annealers using Classical
  Hardness}}}} (\bibinfo{year}{2015}), \bibinfo{note}{(arXiv:1502.02494)}.

\end{thebibliography}

\end{document}